\documentclass[12pt,english,english,12pt,onecolumn, draftcls]{IEEEtran}
\usepackage[T1]{fontenc}
\usepackage[latin9]{inputenc}
\usepackage{textcomp}
\usepackage{amsthm}
\usepackage{amsmath}
\usepackage{amssymb}
\usepackage{graphicx}
\usepackage{esint}
\PassOptionsToPackage{version=3}{mhchem}
\usepackage{mhchem}

\makeatletter
\theoremstyle{plain}
\newtheorem{thm}{\protect\theoremname}
\theoremstyle{definition}
\newtheorem{defn}[thm]{\protect\definitionname}
\theoremstyle{remark}
\newtheorem{rem}[thm]{\protect\remarkname}
\theoremstyle{plain}
\newtheorem{prop}[thm]{\protect\propositionname}
\theoremstyle{plain}
\newtheorem{cor}[thm]{\protect\corollaryname}

\ifCLASSINFOpdf
\else
\fi
\theoremstyle{plain}
\newtheorem{coro}{Corollary}
\renewenvironment{cor}{\begin{coro}}{\end{coro}}
\newtheorem{propo}{Proposition}
\renewenvironment{prop}{\begin{propo}}{\end{propo}}
\theoremstyle{remark}
\newtheorem{rema}{Remark}
\renewenvironment{rem}{\begin{rema}}{\end{rema}}

\usepackage{babel}
\providecommand{\corollaryname}{Corollary}
\providecommand{\definitionname}{Definition}
\providecommand{\propositionname}{Proposition}
\providecommand{\remarkname}{Remark}
\providecommand{\theoremname}{Theorem}

\usepackage{babel}
\providecommand{\corollaryname}{Corollary}
\providecommand{\definitionname}{Definition}
\providecommand{\propositionname}{Proposition}
\providecommand{\remarkname}{Remark}
\providecommand{\theoremname}{Theorem}

\makeatother

\usepackage{babel}
\providecommand{\corollaryname}{Corollary}
\providecommand{\definitionname}{Definition}
\providecommand{\propositionname}{Proposition}
\providecommand{\remarkname}{Remark}
\providecommand{\theoremname}{Theorem}

\begin{document}

\title{Cascade Source Coding with a Side Information {}``Vending Machine''}

\author{Behzad~Ahmadi, Chiranjib Choudhuri, Osvaldo Simeone and Urbashi
Mitra 
\thanks{B. Ahmadi and O. Simeone are with the CWCSPR, New Jersey Institute
of Technology, Newark, NJ 07102 USA (e-mail: \{behzad.ahmadi,osvaldo.simeone\}@njit.edu). %
}%
\thanks{C. Choudhuri and U. Mitra are with Ming Hsieh Dept. of Electrical
Engineering, University of Southern California, Los Angeles, CA, 90089
USA (e-mail: \{cchoudhu,ubli\}@usc.edu).%
}}

\maketitle







\begin{abstract}
The model of a side information {}``vending machine\textquotedblright{}
(VM) accounts for scenarios in which the measurement of side information
sequences can be controlled via the selection of cost-constrained
actions. In this paper, the three-node cascade source coding problem
is studied under the assumption that a side information VM is available
and the intermediate and/or at the end node of the cascade. A single-letter
characterization of the achievable trade-off among the transmission
rates, the distortions in the reconstructions at the intermediate
and at the end node, and the cost for acquiring the side information
is derived for a number of relevant special cases. It is shown that
a joint design of the description of the source and of the control
signals used to guide the selection of the actions at downstream nodes
is generally necessary for an efficient use of the available communication
links. In particular, for all the considered models, layered coding
strategies prove to be optimal, whereby the base layer fulfills two
network objectives: determining the actions of downstream nodes and
simultaneously providing a coarse description of the source. Design
of the optimal coding strategy is shown via examples to depend on
both the network topology and the action costs. Examples also illustrate
the involved performance trade-offs across the network.\end{abstract}
\IEEEpeerreviewmaketitle

\section{Introduction}

The concept of a side information {}``vending machine\textquotedblright{}
(VM) was introduced in \cite{Permuter} for a point-to-point model,
in order to account for source coding scenarios in which acquiring
the side information at the receiver entails some cost and thus should
be done efficiently. In this class of models, the quality of the side
information $Y$ can be controlled at the decoder by selecting an
action $A$ that affects the effective channel between the source
$X$ and the side information $Y$ through a conditional distribution
$p_{Y|X,A}(y|x,a)$. Each action $A$ is associated with a cost, and
the problem is that of characterizing the available trade-offs between
rate, distortion and action cost.

\begin{figure}[h!]
\centering\includegraphics[bb=10bp 10bp 592bp 325bp,clip,scale=0.45]{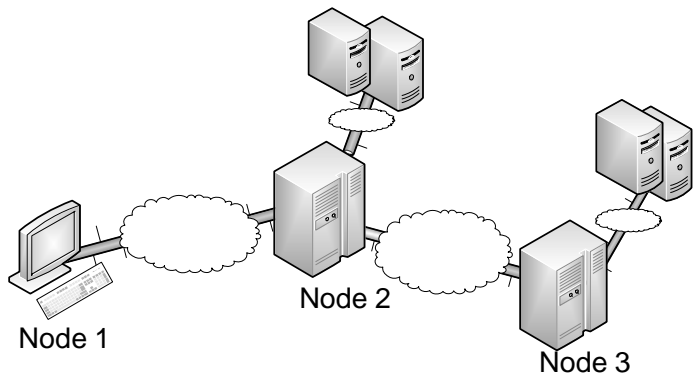}
\caption{A multi-hop computer network in which intermediate and end nodes can
access side information by interrogating remote data bases via cost-constrained
actions.}

\label{fig:fig1} 
\end{figure}

Extending the point-to-point set-up, cascade models provide baseline
scenarios in which to study fundamental aspects of communication in
multi-hop networks, which are central to the operation of, e.g., sensor
or computer networks (see Fig. \ref{fig:fig1}). Standard information-theoretic
models for cascade scenarios assume the availability of given side
information sequences at the nodes (see e.g., \cite{Ravi}-\cite{Chia}).
In this paper, instead, we account for the cost of acquiring the side
information by introducing side information VMs at an intermediate
node and/ or at the final destination of a cascade model. As an example
of the applications of interest, consider the computer network of
Fig. \ref{fig:fig1}, where the intermediate and end nodes can obtain
side information from remote data bases, but only at the cost of investing
system resources such as time or bandwidth. Another example is a sensor
network in which acquiring measurements entails an energy cost.

As shown in \cite{Permuter} for a point-to-point system, the optimal
operation of a VM at the decoder requires taking actions that are
guided by the message received from the encoder. This implies the
exchange of an explicit control signal embedded in the message communicated
to the decoder that instructs the latter on how to operate the VM.
Generalizing to the cascade models under study, a key issue\emph{
}to be tackled in this work is the design of communication strategies
that strike the right balance between control signaling and source
compression across the two hops.

\subsection{Related Work}

As mentioned, the original paper \cite{Permuter} considered a point-to-point
system with a single encoder and a single decoder. Various works have
extended the results in \cite{Permuter} to multi-terminal models.
Specifically, \cite{Weissman_multi,Ahmadi_ISIT'11} considered a set-up
analogous to the Heegard-Berger problem \cite{HB,Kaspi}, in which
the side information may or may not be available at the decoder. The
more general case in which both decoders have access to the same vending
machine, and either the side information produced by the vending machine
at the two decoders satisfy a degradedness condition, or lossless
source reconstructions are required at the decoders is solved in \cite{Weissman_multi}.
In \cite{Ahmadi_DSC}, a distributed source coding setting that extends
\cite{Berger-Yeung} to the case of a decoder with a side information
VM is investigated, along with a cascade source coding model to be
discussed below. Finally, in \cite{Compression with actions}, a related
problem is considered in which the sequence to be compressed is dependent
on the actions taken by a separate encoder.

\begin{figure}[h!]
\centering \includegraphics[bb=67bp 377bp 515bp 706bp,clip,scale=0.65]{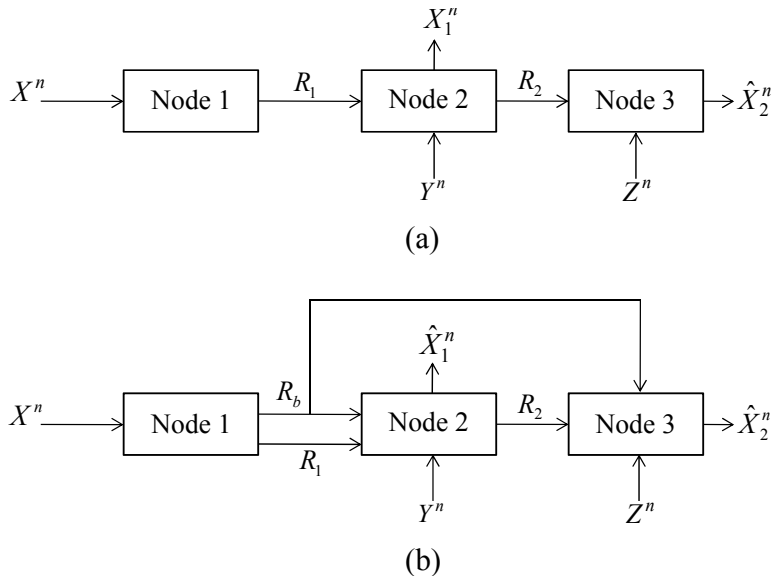}
\caption{($a$) Cascade source coding problem and ($b$) cascade-broadcast
source coding problem.}

\label{fig:intro} 
\end{figure}

The problem of characterizing the rate-distortion region for cascade
source coding models, even with conventional side information sequences
(i.e., without VMs as in Fig. \ref{fig:intro}) at Node 2 and Node
3, is generally open. We refer to \cite{Ravi} and references therein
for a review of the state of the art on the cascade problem and to
\cite{Vasudevan} for the cascade-broadcast problem.

In this work, we focus on the cascade source coding problem with side
information VMs. The basic cascade source coding model consists of
three nodes arranged so that Node 1 communicates with Node 2 and Node
2 to Node 3 over finite-rate links, as illustrated for a computer
network scenario in Fig. \ref{fig:fig1} and schematically in Fig.
\ref{fig:intro}-(a). Both Node 2 and Node 3 wish to reconstruct a,
generally lossy, version of source $X$ and have access to different
side information sequences. An extension of the cascade model is the
cascade-broadcast model of Fig. \ref{fig:intro}-(b), in which an
additional \textquotedbl{}broadcast\textquotedbl{} link of rate $R_{b}$
exists that is received by both Node 2 and Node 3.

Two specific instances of the models in Fig. \ref{fig:intro} for
which a characterization of the rate-distortion performance has been
found are the settings considered in \cite{Chia} and that in \cite{Ahmadi_CR},
which we briefly review here for their relevance to the present work.
In \cite{Chia}, the cascade model in Fig. \ref{fig:intro}(a) was
considered for the special case in which the side information $Y$
measured at Node 2 is also available at Node 1 (i.e., $X=(X,Y)$)
and we have the Markov chain $X-Y-Z$ so that the side information
at Node 3 is degraded with respect to that of Node 2. Instead, in
\cite{Ahmadi_CR}, the cascade-broadcast model in Fig. \ref{fig:intro}(b)
was considered for the special case in which either rate $R_{b}$
or $R_{1}$ is zero, and the reconstructions at Node 1 and Node 2
are constrained to be retrievable also at the encoder in the sense
of the Common Reconstruction (CR) introduced in \cite{Steinberg}
(see below for a rigorous definition).

\subsection{Contributions}

In this paper, we investigate the source coding models of Fig. \ref{fig:intro}
by assuming that some of the side information sequences can be affected
by the actions taken by the corresponding nodes via VMs. The main
contributions are as follows. 
\begin{itemize}
\item \emph{Cascade source coding problem with VM at Node 3} (Fig. \ref{fig:fig2}):
In Sec. \ref{sub:RD_cascade}, we derive the achievable rate-distortion-cost
trade-offs for the set-up in Fig. \ref{fig:fig2}, in which a side
information VM exists at Node 3, while the side information $Y$ is
known at both Node 1 and Node 2 and satisfies the Markov chain $X\textrm{---}Y\textrm{---}Z$.
This characterization extends the result of \cite{Chia} discussed
above to a model with a VM at Node 3. We remark that in \cite{Ahmadi_DSC},
the rate-distortion-cost characterization for the model in Fig. \ref{fig:fig2}
was obtained, but under the assumption that the side information at
Node 3 be available in a causal fashion in the sense of \cite{Weiss-Elgam}; 
\item \emph{Cascade-broadcast source coding problem with VM at Node 2 and
Node 3, lossless compression} (Fig. \ref{fig:fig3}): In Sec. \ref{sub:RD_BC_lossless},
we study the cascade-broadcast model in Fig. \ref{fig:fig3} in which
a VM exists at both Node 2 and Node 3. In order to enable the action
to be taken by both Node 2 and Node 3, we assume that the information
about which action should be taken by Node 2 and Node 3 is sent by
Node 1 on the broadcast link of rate $R_{b}$. Under the constraint
of lossless reconstruction at Node 2 and Node 3, we obtain a characterization
of the rate-cost performance. This conclusion generalizes the result
in \cite{Weissman_multi} discussed above to the case in which the
rate $R_{1}$ and/or $R_{2}$ are non-zero; 
\item \emph{Cascade-broadcast source coding problem with VM at Node 2 and
Node 3, lossy compression with CR constraint} (Fig. \ref{fig:fig3}):
In Sec. \ref{sub:CR}, we tackle the problem in Fig. \ref{fig:fig3}
but under the more general requirement of lossy reconstruction. Conclusive
results are obtained under the additional constraints that the side
information at Node 3 is degraded and that the source reconstructions
at Node 2 and Node 3 can be recovered with arbitrarily small error
probability at Node 1. This is referred to as the CR constraint following
\cite{Steinberg}, and is of relevance in applications in which the
data being sent is of sensitive nature and unknown distortions in
the receivers' reconstructions are not acceptable (see \cite{Steinberg}
for further discussion). This characterization extends the result
of \cite{Ahmadi_CR} mentioned above to the set-up with a side information
VM, and also in that both rates $R_{1}$ and $R_{b}$ are allowed
to be non-zero; 
\item \emph{Adaptive actions}: Finally, we revisit the results above by
allowing the decoders to select their actions in an adaptive way,
based not only on the received messages but also on the previous samples
of the side information extending \cite{Chiru}. Note that the effect
of adaptive actions on rate--distortion--cost region was open even
for simple point-to-point communication channel with decoder side
non-causal side information VM until recently, when \cite{Chiru}
has shown that adaptive action does not decrease the rate--distortion--cost
region of point-to-point system. In this paper we have extended this
result to the multi-terminal framework and we conclude that, in all
of the considered examples, where applicable, adaptive
\begin{figure}[h!]
\centering \includegraphics[bb=69bp 530bp 516bp 704bp,clip,scale=0.65]{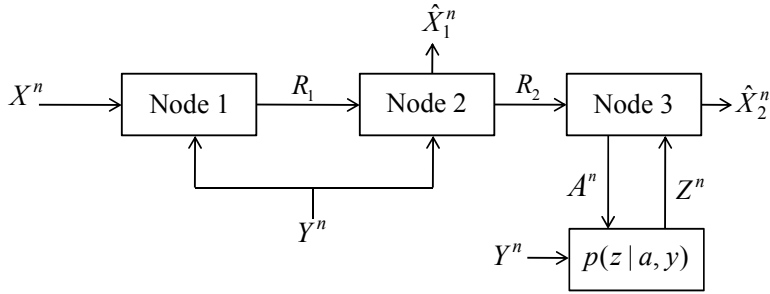}
\caption{Cascade source coding problem with a side information {}``vending
machine'' at Node 3.}

\label{fig:fig2} 
\end{figure}
 selection of the actions does not improve the achievable rate-distortion-cost
trade-offs. 
\end{itemize}
Our results extends to multi-hop scenarios the conclusion in \cite{Permuter}
that a joint representation of data and control messages enables an
efficient use of the available communication links. In particular,
layered coding strategies prove to be optimal for all the considered
models, in which, the base layer fulfills two objectives: determining
the actions of downstream nodes and simultaneously providing a coarse
description of the source. Moreover, the examples provided in the
paper demonstrate the dependence of the optimal coding design on network
topology action costs.

Throughout the paper, we closely follow the notation in \cite{Ahmadi_CR}.
In particular, a random variable is denoted by an upper case letter~(e.g.,
$X,Y,Z$) and its realization is denoted by a lower case letter~(e.g.,
$x,y,z$). The shorthand notation $X^{n}$ is used to denote the tuple
(or the column vector) of random variables $(X_{1},\ldots,X_{n})$,
and $x^{n}$ is used to denote a realization. The notation $X^{n}\sim p(x^{n})$
indicates that $p(x^{n})$ is the probability mass function (pmf)
of the random vector $X^{n}$. Similarly, $Y^{n}|\{X^{n}=x^{n}\}\sim p(y^{n}|x^{n})$
indicates that $p(y^{n}|x^{n})$ is the conditional pmf of $Y^{n}$
given $\{X^{n}=x^{n}\}$. We say that $X\textrm{---}Y\textrm{---}Z$
form a Markov chain if $p(x,y,z)=p(x)p(y|x)p(z|y)$, that is, $X$
and $Z$ are conditionally independent of each other given $Y$.

\section{Cascade Source Coding with A Side information Vending Machine}

In this section, we first describe the system model for the cascade
source coding problem with a side information vending machine of Fig.
\ref{fig:fig2}. We then present the characterization of the corresponding
rate-distortion-cost performance in Sec. \ref{sub:RD_cascade}. 
\begin{figure}[h!]
\centering \includegraphics[bb=67bp 490bp 515bp 692bp,clip,scale=0.65]{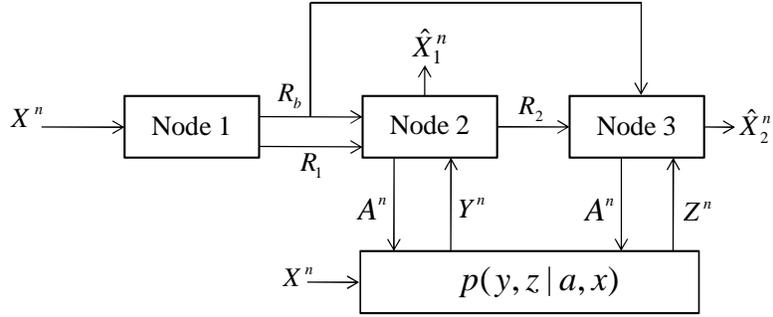}
\caption{Cascade source coding problem with a side information {}``vending
machine'' at Node 2 and Node 3.}

\label{fig:fig3} 
\end{figure}

\subsection{System Model\label{sub:System-Model_cascade}}

The problem of cascade source coding of Fig. \ref{fig:fig2}, is defined
by the probability mass functions (pmfs) $p_{XY}(x,y)$ and $p_{Z|AY}(z|a,y)$
and discrete alphabets $\mathcal{X},\mathcal{Y},\mathcal{Z},\mathcal{A},\mathcal{\hat{X}}_{1},\mathcal{\hat{X}}_{2},$
as follows. The source sequences $X^{n}$ and $Y^{n}$ with $X^{n}\in\mathcal{X}^{n}$
and $Y^{n}\in\mathcal{Y}^{n}$, respectively$,$ are such that the
pairs $(X_{i},Y_{i})$ for $i\in[1,n]$ are independent and identically
distributed (i.i.d.) with joint pmf $p_{XY}(x,y)$. Node 1 measures
sequences $X^{n}$ and $Y^{n}$ and encodes them in a message $M_{1}$
of $nR_{1}$ bits, which is delivered to Node 2. Node 2 estimates
a sequence $\hat{X}_{1}^{n}\in\mathcal{\hat{X}}_{1}^{n}$ within given
distortion requirements to be discussed below. Moreover, Node 2 maps
the message $M_{1}$ received from Node 1 and the locally available
sequence $Y^{n}$ in a message $M_{2}$ of $nR_{2}$ bits, which is
delivered to Node 3. Node 3 wishes to estimate a sequence $\hat{X}_{2}^{n}\in\mathcal{\hat{X}}_{2}^{n}$
within given distortion requirements. To this end, Node 3 receives
message $M_{2}$ and based on this, it selects an action sequence
$A^{n},$ where $A^{n}\in\mathcal{A}^{n}.$ The action sequence affects
the quality of the measurement $Z^{n}$ of sequence $Y^{n}$ obtained
at the Node 3. Specifically, given $A^{n}$ and $Y^{n}$, the sequence
$Z^{n}$ is distributed as $p(z^{n}|a^{n},y^{n})=\prod_{i=1}^{n}p_{Z|A,Y}(z_{i}|y_{i},a_{i})$.
The cost of the action sequence is defined by a cost function $\Lambda$:
$\mathcal{A\rightarrow}[0,\Lambda_{\max}]$ with $0\leq\Lambda_{\max}<\infty,$
as $\Lambda(a^{n})=\sum_{i=1}^{n}\Lambda(a_{i})$. The estimated sequence
$\hat{X}_{2}^{n}$ with $\hat{X}_{2}^{n}\in\mathcal{\hat{X}}_{2}^{n}$
is then obtained as a function of $M_{2}$ and $Z^{n}$. The estimated
sequences $\hat{X}_{j}^{n}$ for $j=1,2$ must satisfy distortion
constraints defined by functions $d_{j}(x,\hat{x}_{j})$: $\mathcal{X}\times\mathcal{\hat{X}}_{j}\rightarrow[0,D_{\max}]$
with $0\leq D_{\max}<\infty$ for $j=1,2,$ respectively. A formal
description of the operations at the encoder and the decoder follows.
\begin{figure}[h!]
\centering \includegraphics[bb=60bp 473bp 520bp 679bp,clip,scale=0.72]{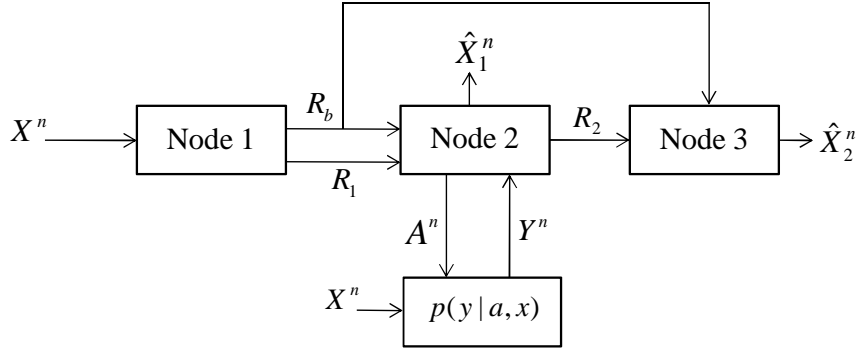}
\caption{Cascade-broadcast source coding problem with a side information {}``vending
machine'' at Node 2.}

\label{fig:fig4} 
\end{figure}

\begin{defn}
\label{def_cascade}An $(n,R_{1},R_{2},D_{1},D_{2},\Gamma,\epsilon)$
code for the set-up of Fig. \ref{fig:fig2} consists of two source
encoders, namely 
\begin{equation}
\mathrm{g}_{1}\text{: }\mathcal{X}^{n}\times\mathcal{Y}^{n}\rightarrow[1,2^{nR_{1}}],\label{encoder1}
\end{equation}
which maps the sequences $X^{n}$ and $Y^{n}$ into a message $M_{1};$
\begin{equation}
\mathrm{g}_{2}\text{:}\text{ }\mathcal{Y}^{n}\times[1,2^{nR_{1}}]\rightarrow[1,2^{nR_{2}}],\label{encoder2}
\end{equation}
which maps the sequence $Y^{n}$ and message $M_{1}$ into a message
$M_{2};$ an {}``action\textquotedblright{}\ function 
\begin{equation}
\mathrm{\ell}\text{: }[1,2^{nR_{2}}]\rightarrow\mathcal{A}^{n},\label{action_fun}
\end{equation}
which maps the message $M_{2}$ into an action sequence $A^{n};$
two decoders, namely 
\begin{equation}
\mathrm{h}_{1}\text{: }[1,2^{nR_{1}}]\times\mathcal{Y}^{n}\rightarrow\mathcal{\hat{X}}_{1}^{n},\label{decoder1}
\end{equation}
which maps the message $M_{1}$ and the measured sequence $Y^{n}$
into the estimated sequence $\hat{X}_{1}^{n};$ 
\begin{equation}
\mathrm{h}_{2}\text{: }[1,2^{nR_{2}}]\times\mathcal{Z}^{n}\rightarrow\mathcal{\hat{X}}_{2}^{n},\label{decoder2}
\end{equation}
which maps the message $M_{2}$ and the measured sequence $Z^{n}$
into the the estimated sequence $\hat{X}_{2}^{n};$ such that the
action cost constraint $\Gamma$ and distortion constraints $D_{j}$
for $j=1,2$ are satisfied, i.e., 
\begin{align}
\frac{1}{n}\underset{i=1}{\overset{n}{\sum}}\mathrm{E}\left[\Lambda(A_{i})\right] & \leq\Gamma\label{action cost}\\
\text{ and }\frac{1}{n}\underset{i=1}{\overset{n}{\sum}}\mathrm{E}\left[d_{j}(X_{ji},\textrm{h}_{ji})\right] & \leq D_{j}\text{ for }j=1,2,\label{dist const}
\end{align}
where we have defined as $\textrm{h}_{1i}$ and $\textrm{h}_{2i}$
the $i$th symbol of the function $\textrm{h}_{1}(M_{1},Y^{n})$ and
$\textrm{h}_{2}(M_{2},Z^{n})$, respectively. 
\end{defn}

\begin{defn}
\label{def_ach}Given a distortion-cost tuple $(D_{1},D_{2},\Gamma)$,
a rate tuple $(R_{1},R_{2})$ is said to be achievable if, for any
$\epsilon>0$, and sufficiently large $n$, there exists a $(n,R_{1},R_{2},D_{1}+\epsilon,D_{2}+\epsilon,\Gamma+\epsilon)$
code. 
\end{defn}

\begin{defn}
\label{def_reg}The \textit{rate-distortion-cost region }$\mathcal{R}(D_{1},D_{2},\Gamma)$
is defined as the closure of all rate tuples $(R_{1},R_{2})$ that
are achievable given the distortion-cost tuple $(D_{1},D_{2},\Gamma)$. \end{defn}
\begin{rem}
For side information $Z$ available causally at Node 3, i.e., with
decoding function (\ref{decoder2}) at Node 3 modified so that $\hat{X}_{i}$
is a function of $M_{2}$ and $Z^{i}$ only, the rate-distortion region
$\mathcal{R}(D_{1},D_{2},\Gamma)$ has been derived in \cite{Ahmadi_DSC}. 
\end{rem}
In the rest of this section, for simplicity of notation, we drop the
subscripts from the definition of the pmfs, thus identifying a pmf
by its argument.

\subsection{Rate-Distortion-Cost Region \label{sub:RD_cascade}}

In this section, a single-letter characterization of the rate-distortion-cost
region is derived. 
\begin{prop}
\label{prop:RD_action_cascade}The rate-distortion-cost region $\mathcal{R\mbox{\ensuremath{(D_{1},D_{2},\Gamma)}}}$
for the cascade source coding problem illustrated in Fig. \ref{fig:fig2}
is given by the union of all rate pairs $(R_{1},R_{2})$ that satisfy
the conditions\begin{subequations}\label{eqn: RD_action_cascade}
\begin{eqnarray}
R_{1} & \geq & I(X;\hat{X}_{1},A,U|Y)\label{eq:R1}\\
\ce{and}\mbox{ }R_{2} & \geq & I(X,Y;A)+I(X,Y;U|A,Z),\label{eq:R2}
\end{eqnarray}
\end{subequations}where the mutual information terms are evaluated
with respect to the joint pmf 
\begin{align}
p(x,y,z,a,\hat{x}_{1},u)=p(x,y)p(\hat{x}_{1},a,u|x,y)p(z|y,a) & ,\label{eq:joint}
\end{align}
for some pmf $p(\hat{x}_{1},a,u|x,y)$ such that the inequalities\begin{subequations}\label{eqn: action_cascade_const}
\begin{eqnarray}
\ce{E}[d_{1}(X,\hat{X}_{1})] & \leq & D_{1},\label{eq:dist1}\\
\ce{E}[d_{2}(X,\ce{f}(U,Z))] & \leq & D_{2},\label{eq:dist2}\\
\ce{and}\textrm{ }\ce{E}[\Lambda(A)] & \leq & \Gamma,\label{eq:action_bound}
\end{eqnarray}
\end{subequations}are satisfied for some function $\ce{f}\textrm{: }\mathcal{U}\times\mathcal{Z}\rightarrow\hat{\mathcal{X}}_{2}$.
Finally, $U$ is an auxiliary random variable whose alphabet cardinality
can be constrained as $|\mathcal{U}|\leq|\mathcal{X}||\mathcal{Y}||\mathcal{A}|+3$,
without loss of optimality. \end{prop}
\begin{rem}
For side information $Z$ independent of the action $A$ given $Y$$,$
i.e., for $p(z|a,y)=p(z|y),$ the rate-distortion region $\mathcal{R}(D_{1},D_{2},\Gamma)$
in Proposition \ref{prop:RD_action_cascade} reduces to that derived
in \cite{Chia}. 
\end{rem}
The proof of the converse is provided in Appendix A for a more general
case of adaptive action to be defined in Sec \ref{sec:Adaptive-Actions}.
The achievability follows as a combination of the techniques proposed
in \cite{Permuter} and \cite[Theorem 1]{Chia}. Here we briefly outline
the main ideas, since the technical details follow from standard arguments.
For the scheme at hand, Node 1 first maps sequences $X^{n}$ and $Y^{n}$
into the action sequence $A^{n}$ using the standard joint typicality
criterion. This mapping requires a codebook of rate $I(X,Y;A)$ (see,
e.g., \cite[pp. 62-63]{Elgammal})$.$ Given the sequence $A^{n}$,
the sequences $X^{n}$ and $Y^{n}$ are further mapped into a sequence
$U^{n}$. This requires a codebook of size $I(X,Y;U|A)$ for each
action sequence $A^{n}$ from standard rate-distortion considerations
\cite[pp. 62-63]{Elgammal}. Similarly, given the sequences $A^{n}$
and $U^{n},$ the sequences $X^{n}$ and $Y^{n}$ are further mapped
into the estimate $\hat{X}_{1}^{n}$ for Node 2 using a codebook of
rate $I(X,Y;\hat{X}_{1}|U,A)$ for each codeword pair $(U^{n},A^{n})$.
The thus obtained codewords are then communicated to Node 2 and Node
3 as follows. By leveraging the side information $Y^{n}$ available
at Node 2, conveying the codewords $A^{n},$ $U^{n}$ and $\hat{X}_{1}^{n}$
to Node 2 requires rate $I(X,Y;U,A)+I(X,Y;\hat{X}_{1}|U,A)-I(U,A,\hat{X}_{1};Y)$
by the Wyner-Ziv theorem \cite[p. 280]{Elgammal}$,$ which equals
the right-hand side of (\ref{eq:R1}). Then, sequences $A^{n}$ and
$U^{n}$ are sent by Node 2 to Node 3, which requires a rate equal
to the right-hand side of (\ref{eq:R2}). This follows from the rates
of the used codebooks and from the Wyner-Ziv theorem, due to the side
information $Z^{n}$ available at Node 3 upon application of the action
sequence $A^{n}$. Finally, Node 3 produces $\hat{X}_{2}^{n}$ that
leverages through a symbol-by-symbol function as $\hat{X}_{2i}=\textrm{f}(U_{i},Z_{i})$
for $i\in[1,n].$

\subsection{Lossless Compression}

Suppose that the source sequence $X^{n}$ needs to be communicated
{\em losslessly} at both Node 2 and Node 3, in the sense that $d_{j}(x,\hat{x}_{j})$
is the Hamming distortion measure for $j=1,2$ ($d_{j}(x,\hat{x}_{j})=0$
if $x=\hat{x}_{j}$ and $d_{j}(x,\hat{x}_{j})=1$ if $x\neq\hat{x}_{j}$)
and $D_{1}=D_{2}=0$. We can establish the following immediate consequence
of Proposition~\ref{prop:RD_action_cascade}. \begin{coro}\label{coro:cascade_lossless}
The rate-distortion-cost region $\mathcal{R\mbox{\ensuremath{(0,0,\Gamma)}}}$
for the cascade source coding problem illustrated in Fig. \ref{fig:fig2}
with Hamming distortion metrics is given by the union of all rate
pairs $(R_{1},R_{2})$ that satisfy the conditions\begin{subequations}\label{eqn: R_action_cascade_lossless}
\begin{eqnarray}
R_{1} & \geq & I(X;A|Y)+H(X|A,Y)\label{eq:R1_lossless}\\
\ce{and}\mbox{ }R_{2} & \geq & I(X,Y;A)+H(X|A,Z),\label{eq:R2_lossless}
\end{eqnarray}
\end{subequations}where the mutual information terms are evaluated
with respect to the joint pmf 
\begin{align}
p(x,y,z,a)=p(x,y)p(a|x,y)p(z|y,a) & ,\label{eq:joint_lossless}
\end{align}
for some pmf $p(a|x,y)$ such that $\ce{E}[\Lambda(A)]\leq\Gamma$.
\end{coro}

\section{Cascade-Broadcast Source Coding with A Side Information Vending Machine}

In this section, the cascade-broadcast source coding problem with
a side information vending machine illustrated in Fig. \ref{fig:fig3}
is studied. At first, the rate-cost performance is characterized for
the special case in which the reproductions at Node 2 and Node 3 are
constrained to be lossless. Then, the lossy version of the problem
is considered in Sec. \ref{sub:CR}, with an additional common reconstruction
requirement in the sense of \cite{Steinberg} and assuming degradedness
of the side information sequences.

\subsection{System Model\label{sub:lossless}}

In this section, we describe the general system model for the cascade-broadcast
source coding problem with a side information vending machine. We
emphasize that, unlike the setup of Fig. \ref{fig:fig2}, here, the
vending machine is at both Node 2 and Node 3. Moreover, we assume
that an additional broadcast link of rate $R_{b}$ is available that
is received by Node 2 and 3 to enable both Node 2 and Node 3 so as
to take concerted actions in order to affect the side information
sequences. We assume the action sequence taken by Node 2 and Node
3 to be a function of only the broadcast message $M_{b}$ sent over
the broadcast link of rate $R_{b}$.

The problem is defined by the pmfs $p_{X}(x)$, $p_{YZ|AX}(y,z|a,x)$
and discrete alphabets $\mathcal{X},\mathcal{Y},{\cal Z},\mathcal{A},$
$\mathcal{\hat{X}}_{1},\mathcal{\hat{X}}_{2},$ as follows. The source
sequence $X^{n}$ with $X^{n}\in\mathcal{X}^{n}$ is i.i.d. with pmf
$p_{X}(x)$. Node 1 measures sequence $X^{n}$ and encodes it into
messages $M_{1}$ and $M_{b}$ of $nR_{1}$ and $nR_{b}$ bits, respectively,
which are delivered to Node 2. Moreover, message $M_{b}$ is broadcast
also to Node 3. Node 2 estimates a sequence $\hat{X}_{1}^{n}\in\mathcal{\hat{X}}_{1}^{n}$
and Node 3 estimates a sequence $\hat{X}_{2}^{n}\in\mathcal{\hat{X}}_{2}^{n}$.
To this end, Node 2 receives messages $M_{1}$ and $M_{b}$ and, based
only on the latter message, it selects an action sequence $A^{n},$
where $A^{n}\in\mathcal{A}^{n}.$ Node 2 maps messages $M_{1}$ and
$M_{b}$, received from Node 1, and the locally available sequence
$Y^{n}$ in a message $M_{2}$ of $nR_{2}$ bits, which is delivered
to Node 3. Node 3 receives messages $M_{2}$ and $M_{b}$ and based
only on the latter message, it selects an action sequence $A^{n},$
where $A^{n}\in\mathcal{A}^{n}.$ Given $A^{n}$ and $X^{n}$, the
sequences $Y^{n}$ and $Z^{n}$ are distributed as $p(y^{n},z^{n}|a^{n},x^{n})=\prod_{i=1}^{n}p_{YZ|A,X}(y_{i},z_{i}|a_{i},x_{i})$.
The cost of the action sequence is defined as in previous section.
A formal description of the operations at encoder and decoder follows. 
\begin{defn}
\label{def_BC_lossless}An $(n,R_{1},R_{2},R_{b},D_{1},D_{2},\Gamma,\epsilon)$
code for the set-up of Fig. \ref{fig:fig4} consists of two source
encoders, namely 
\begin{eqnarray}
\mathrm{g}_{1}\text{: \ensuremath{\mathcal{X}^{n}\rightarrow[1,2^{nR_{1}}]\times[1,2^{nR_{b}}],}} &  & \text{ }\label{eq:encoder1}
\end{eqnarray}
which maps the sequence $X^{n}$ into messages $M_{1}$ and $M_{b}$,
respectively; 
\begin{equation}
\mathrm{g}_{2}\text{:}\text{ }[1,2^{nR_{1}}]\times[1,2^{nR_{b}}]\times\mathcal{Y}^{n}\rightarrow[1,2^{nR_{2}}]\label{encoder2_BC}
\end{equation}
which maps the sequence $Y^{n}$ and messages $(M_{1},M_{b})$ into
a message $M_{2};$ an {}``action\textquotedblright{}\ function
\begin{equation}
\mathrm{\ell}\text{: }[1,2^{nR_{b}}]\rightarrow\mathcal{A}^{n},\label{action_fun_BC}
\end{equation}
which maps the message $M_{b}$ into an action sequence $A^{n};$
two decoders, namely 
\begin{equation}
\mathrm{h}_{1}\text{: }\text{ }[1,2^{nR_{1}}]\times[1,2^{nR_{b}}]\times\mathcal{Y}^{n}\rightarrow\mathcal{\hat{X}}_{1}^{n},\label{decoder1_BC}
\end{equation}
which maps messages $M_{1}$ and $M_{b}$ and the measured sequence
$Y^{n}$ into the estimated sequence $\hat{X}_{1}^{n};$ and 
\begin{equation}
\mathrm{h}_{2}\text{: }[1,2^{nR_{2}}]\times[1,2^{nR_{b}}]\times\mathcal{Z}^{n}\rightarrow\mathcal{\hat{X}}_{2}^{n},\label{decoder2_BC}
\end{equation}
which maps the messages $M_{2}$ and $M_{b}$ into the the estimated
sequence $\hat{X}_{2}^{n};$ such that the action cost constraint
(\ref{action cost}) and distortion constraint (\ref{dist const})
are satisfied.

Achievable rates $(R_{1},R_{2},R_{b})$ and rate-distortion-cost region
are defined analogously to Definitions \ref{def_ach} and \ref{def_reg}. 
\end{defn}
The rate--distortion--cost region for the system model described above
is open even for the case without VM at Node 2 and Node 3 (see \cite{Vasudevan}).
Hence, in the following subsections, we characterize the rate region
for a few special cases. As in the previous section, subscripts are
dropped from the pmf for simplicity of notation.

\subsection{Lossless Compression\label{sub:RD_BC_lossless}}

In this section, a single-letter characterization of the rate-cost
region $\mathcal{R\mbox{\ensuremath{(0,0,\Gamma)}}}$ is derived for
the special case in which the distortion metrics are assumed to be
Hamming and the distortion constraints are $D_{1}=0$ and $D_{2}=0$. 
\begin{prop}
\label{prop:RD_BC_lossless}The rate-cost region $\mathcal{R\mbox{\ensuremath{(0,0,\Gamma)}}}$
for the cascade-broadcast source coding problem illustrated in Fig.
\ref{fig:fig3} with Hamming distortion metrics is given by the union
of all rate triples $(R_{1},R_{2},R_{b})$ that satisfy the conditions\begin{subequations}\label{eqn: RD_BC_lossless}
\begin{eqnarray}
R_{b} & \geq & I(X;A)\label{eq:Ra_lossless}\\
R_{1}+R_{b} & \geq & I(X;A)+H(X|A,Y)\label{eq:R1+Rb_lossless}\\
\ce{and}\mbox{ }R_{2}+R_{b} & \geq & I(X;A)+H(X|A,Z)\label{eq:R2+Rb_lossless}
\end{eqnarray}
\end{subequations}where the mutual information terms are evaluated
with respect to the joint pmf 
\begin{align}
p(x,y,z,a)=p(x,a)p(y,z|a,x),\label{eq:joint-BC_lossless}
\end{align}
for some pmf $p(a|x)$ such that $\textrm{E}[\Lambda(A)]\leq\Gamma$. \end{prop}
\begin{rem}
If $R_{1}=0$ and $R_{2}=0,$ the rate-cost region $\mathcal{R}(\Gamma)$
of Proposition \ref{prop:RD_BC_lossless} reduces to the one derived
in \cite[Theorem 1]{Weissman_multi}. 
\end{rem}

\begin{rem}
The rate region (\ref{eqn: RD_BC_lossless}) also describes the rate-distortion
region under the more restrictive requirement of lossless reconstruction
in the sense of the probabilities of error $\textrm{Pr}[X^{n}\neq\hat{X}_{j}^{n}]\leq\epsilon$
for $j=1,2$, as it follows from standard arguments (see \cite[Sec. 3.6.4]{Elgammal}).
A similar conclusion applies for Corollary \ref{coro:cascade_lossless}.
\end{rem}
The converse proof for bound (\ref{eq:Ra_lossless}) follows immediately
since $A^{n}$ is selected only as a function of message $M_{b}$.
As for the other two bounds, namely (\ref{eq:R1+Rb_lossless})-(\ref{eq:R2+Rb_lossless}),
the proof of the converse can be established following cut-set arguments
and using the point-to-point result of \cite{Permuter}. For achievability,
we use the code structure proposed in \cite{Permuter} along with
rate splitting. Specifically, Node 1 first maps sequence $X^{n}$
into the action sequence $A^{n}$. This mapping requires a codebook
of rate $I(X;A)$. This rate has to be conveyed over link $R_{b}$
by the definition of the problem and is thus received by both Node
2 and Node 3. Given the so obtained sequence $A^{n}$, communicating
$X$ losslessly to Node 2 requires rate $H(X|A,Y)$. We split this
rate into two rates $r_{1b}$ and $r_{1d}$, such that the message
corresponding to the first rate is carried over the broadcast link
of rate $R_{b}$ and the second on the direct link of rate $R_{1}$.
Note that Node 2 can thus recover sequence $X$ losslessly. The rate
$H(X|A,Z)$ which is required to send $X$ losslessly to Node 3, is
then split into two parts, of rates $r_{2b}$ and $r_{2d}$. The message
corresponding to the rate $r_{2b}$ is sent to Node 3 on the broadcast
link of the rate $R_{b}$ by Node 1, while the message of rate $r_{2d}$
is sent by Node 2 to Node 3. This way, Node 1 and Node 2 cooperate
to transmit $X$ to Node 3. As per the discussion above, the following
inequalities have to be satisfied 
\begin{eqnarray*}
r_{2b}+r_{2d}+r_{1b} & \geq & H(X|A,Z),\\
r_{1b}+r_{1d} & \geq & H(X|A,Y),\\
R_{1} & \geq & r_{1d},\\
R_{2} & \geq & r_{2d},\\
\textrm{and }R_{b} & \geq & r_{1b}+r_{2b}+I(X;A),
\end{eqnarray*}
Applying Fourier-Motzkin elimination \cite[Appendix C]{Elgammal}
to the inequalities above, the inequalities in (\ref{eqn: RD_BC_lossless})
are obtained.

\subsection{Example: Switching-Dependent Side Information\label{sub:lossless_ex}}

We now consider the special case of the model in Fig. \ref{fig:fig3}
in which the actions $A\in{\cal A}=\{0,1,2,3\}$ acts a switch that
decides whether Node 2, Node 3 or either node gets to observe a side
information $W$. The side information $W$ is jointly distributed
with source $X$ according to the joint pmf $p(x,w)$. Moreover, defining
as $\textrm{e}$ an \textquotedbl{}erasure\textquotedbl{} symbol,
the conditional pmf $p(y,z|x,a)$ is as follows: $Y=Z=\textrm{e}$
for $A=0$ (neither Node 2 nor Node 3 observes the side information
$W$); $Y=W$ and $Z=\textrm{e}$ for $A=1$ (only Node 2 observes
the side information $W$); $Y=\textrm{e}$ and $Z=W$ for $A=2$
(only Node 3 observes the side information $W$); and $Y=Z=W$ for
$A=3$ (both nodes observe the side information $W$)%
\footnote{This implies that $p(y,z|x,a)=\underset{w}{\sum}p(w|x)\delta(y-w)\delta(z-\ce{e})$
for $a=1$ and similarly for other values of $a$.%
}. We also select the cost function such that $\Lambda(j)=\lambda_{j}$
for $j\in{\cal A}$. When $R_{1}=R_{2}=0$, this model reduces to
the ones studied in \cite[Sec. III]{Weissman_multi}. The following
is a consequence of Proposition 2. 
\begin{cor}
\label{cor:ex}For the setting of switching-dependent side information
described above, the rate-cost region (\ref{eqn: RD_BC_lossless})
is given by\begin{subequations}\label{eqn: RD_BC_lossless_ex} 
\begin{eqnarray}
R_{b} & \geq & I(X;A)\label{eq:Ra_lossless_ex}\\
R_{1}+R_{b} & \geq & H(X)-p_{1}I(X;W|A=1)-p_{3}I(X;W|A=3)\label{eq:R1+Rb_lossless_ex}\\
\ce{and}\mbox{ }R_{2}+R_{b} & \geq & H(X)-p_{2}I(X;W|A=2)-p_{3}I(X;W|A=3)\label{eq:R2+Rb_lossless_ex}
\end{eqnarray}
\end{subequations}where the mutual information terms are evaluated
with respect to the joint pmf 
\begin{align}
p(x,y,z,a)=p(x,a)p(y,z|a,x),\label{eq:joint-BC_lossless-1}
\end{align}
for some pmf $p(a|x)$ such that $\sum_{j=0}^{3}p_{j}\lambda_{j}\leq\Gamma$,
where we have denoted $p_{j}=\ce{Pr}[A=j]$ for $j\in{\cal A}$.\end{cor}
\begin{IEEEproof}
The region (\ref{eqn: RD_BC_lossless_ex}) is obtained from the rate-cost
region (\ref{eqn: RD_BC_lossless}) by noting that in (\ref{eq:R1+Rb_lossless})
we have $I(X;A)+H(X|A,Y)=H(X)-I(X;Y|A)$ and similarly for (\ref{eq:R2+Rb_lossless}). 
\end{IEEEproof}
In the following, we will elaborate upon two specific instances of
the switching-dependent side information example.

\textit{Binary Symmetric Channel (BSC) between $X$ and $W$}: Let
$(X,W)$ be binary and symmetric so that $p(x)=p(w)=1/2$ for $x,w\in\{0,1\}$
and $\Pr[X\neq W]=\delta$ for $\delta\in[0,1]$. Moreover, let $\lambda_{j}=\infty$
for $j=0,3$ and $\lambda_{j}=1$ otherwise. We set the action cost
constraint to $\Gamma=1$. Note that, given this definition of $\Lambda(a)$,
at each time, Node 1 can choose whether to provide the side information
$W$ to Node 2 \emph{or} to Node 3 with no further constraints. By
symmetry, it can be seen that we can set the pmf $p(a|x)$ with $x\in\{0,1\}$
and $a\in\{1,2\}$ to be a BSC with transition probability $q$. This
implies that $p_{1}=\textrm{Pr}[A=1]=q$ and $p_{2}=\textrm{Pr}[A=2]=1-q$.
We now evaluate the inequality (\ref{eq:Ra_lossless_ex}) as $R_{b}\geq0$;
inequality (\ref{eq:R1+Rb_lossless_ex}) as $R_{1}+R_{b}\geq1-p_{1}I(X;W|A=1)=1-qH(\delta)$;
and similarly inequality (\ref{eq:R2+Rb_lossless}) as $R_{2}+R_{b}\geq1-(1-q)H(\delta).$
From these inequalities, it can be seen that, in order to trace the
boundary of the rate-cost region, in general, one needs to consider
all values of $q$ in the interval $[0,1]$. This
\begin{figure}[h!]
\centering \includegraphics[bb=242bp 431bp 397bp 571bp,clip,scale=0.55]{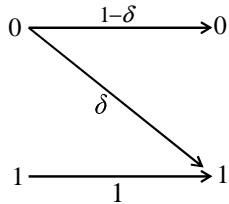}
\caption{The side information S-channel $p(w|x)$ used in the example of Sec.
\ref{sub:lossless_ex}.}

\label{fig:S-Channel} 
\end{figure}
corresponds to appropriate time-sharing between providing side information
to Node 2 (for a fraction of time $q$) and Node 3 (for the remaining
fraction of time). Note that, as shown in \cite[Sec. III]{Weissman_multi},
if $R_{1}=R_{2}=0$, it is optimal to set $q=\frac{1}{2}$, and thus
equally share the side information between Node 2 and Node 3, in order
to minimize the rate $R_{b}$. This difference is due to the fact
that in the cascade model at hand, it can be advantageous to provide
more side information to one of the two encoders depending on the
desired trade-off between the rates $R_{1}$ and $R_{2}$ in the achievable
rate-cost region.

\textit{\emph{S}}\textit{-Channel between $X$ and $W$}: \label{S-Channel}We
now consider the special case of Corollary \ref{cor:ex} in which
$(X,W)$ are jointly distributed so that $p(x)=1/2$ and $p(w|x)$
is the S-channel characterized by $p(0|0)=1-\delta$ and $p(1|1)=1$
(see Fig. \ref{fig:S-Channel}). Moreover, we let $\lambda_{1}=1$,
$\lambda_{2}=0$, $\lambda_{0}=\lambda_{3}=\infty$ as above, while
the cost constraint is set to $\Gamma\leq1$. As discussed in \cite[Sec. III]{Weissman_multi}
for this example with $R_{1}=R_{2}=0$, providing side information
to Node 2 is more costly and thus should be done efficiently. In particular,
given Fig. \ref{fig:S-Channel}, it is expected that biasing the choice
$A=2$ when $X=1$ (i.e., providing side information to Node 2) may
lead to some gain (see \cite{Weissman_multi}). Here we show that
in the cascade model, this gain depends on the relative importance
of rates $R_{1}$ and $R_{2}$. 

To this end, we set $p(a|x)$ as $p(1|0)=\alpha$ and $p(1|1)=\beta$
for $\alpha,\beta\in[0,1]$. We now evaluate the inequality (\ref{eq:Ra_lossless_ex})
as $R_{b}\geq0$; inequality (\ref{eq:R1+Rb_lossless_ex}) as 
\begin{eqnarray}
R_{1}+R_{b} & \geq & 1-\Bigl(\frac{\alpha+\beta}{2}\Bigr)\Bigl(H\Bigl(\frac{(1-\delta)\alpha}{\alpha+\beta}\Bigr)-H(1-\delta)\frac{\alpha}{\alpha+\beta}\Bigr);\label{eq:ex_R1+Rb_ex_S}
\end{eqnarray}
and inequality (\ref{eq:R2+Rb_lossless_ex}) as 
\begin{eqnarray}
R_{2}+R_{b} & \geq & 1-\Bigl(\frac{2-\alpha-\beta}{2}\Bigr)\Bigl(H\Bigl(\frac{(1-\delta)(1-\alpha)}{2-\alpha-\beta}\Bigr)-H(1-\delta)\frac{1-\alpha}{2-\alpha-\beta}\Bigr),\label{eq:ex_R2+Rb_ex_S}
\end{eqnarray}
\begin{figure}[h!]
\centering \includegraphics[bb=50bp 100bp 555bp 550bp,clip,scale=0.6]{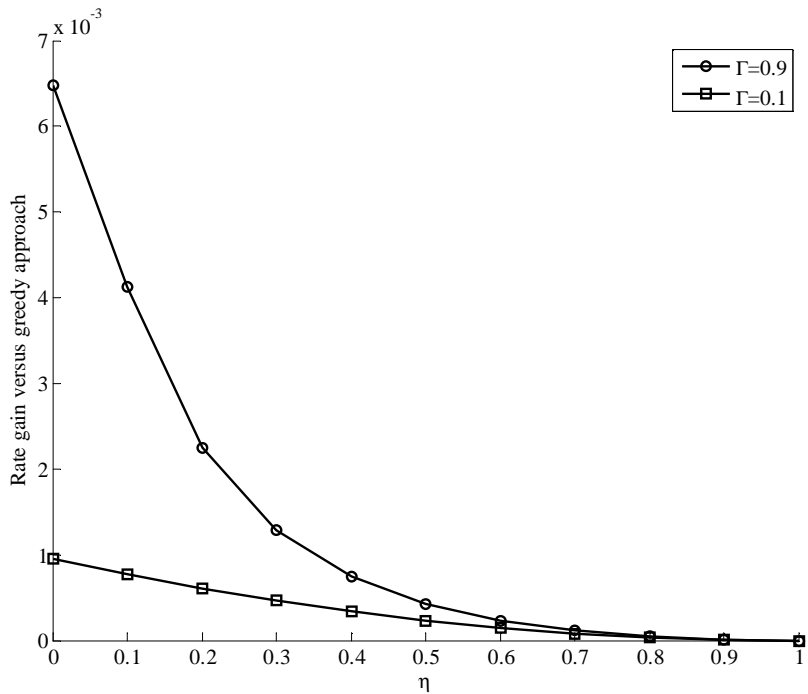}
\caption{Difference between the weighted sum-rate $R_{1}+\eta R_{2}$ obtained
with the greedy and with the optimal strategy as per Corollary \ref{cor:ex}
($R_{b}=0.4$, $\delta=0.6$). }

\label{fig:greedy-nongreedy} 
\end{figure}

We now evaluate the minimum weighted sum-rate $R_{1}+\eta R_{2}$
obtained from (\ref{eq:ex_R1+Rb_ex_S})-(\ref{eq:ex_R2+Rb_ex_S})
for $R_{b}=0.4$, $\delta=0.6$ and both $\Gamma=0.1$ and $\Gamma=0.9$.
Parameter $\eta\geq0$ rules on the relative importance of the two
rates. For comparison, we also compute the performance attainable
by imposing that the action $A$ be selected independent of $X$,
which we refer to as the greedy approach \cite{Permuter}. Fig. \ref{fig:greedy-nongreedy}
plots the difference between the two weighted sum-rates $R_{1}+\eta R_{2}$
. It can be seen that, as $\eta$ decreases and thus minimizing rate
$R_{1}$ to Node 2 becomes more important, one can achieve larger
gains by choosing the action $A$ to be dependent on $X$. Moreover,
this gain is more significant when the action cost budget $\Gamma$
allows Node 2 to collect a larger fraction of the side information
samples.

\subsection{Lossy Compression with Common Reconstruction Constraint\label{sub:CR}}

In this section, we turn to the problem of characterizing the rate-distortion-cost
region ${\cal R}(D_{1},D_{2}$ $,\Gamma)$ for $D_{1},D_{2}>0$. In
order to make the problem tractable %
\footnote{As noted earlier, the problem is open even in the case with no VM
\cite{Vasudevan}.%
}, we impose the degradedness condition $X-(A,Y)-Z$ (as in \cite{Weissman_multi}),
which implies the factorization 
\begin{eqnarray}
p(y,z|a,x) & = & p(y|a,x)p(z|y,a);\label{eq:degradedness}
\end{eqnarray}
and that the reconstructions at Nodes 2 and 3 be reproducible by Node
1. As discussed, this latter condition is referred to as the CR constraint
\cite{Steinberg}. Note that this constraint is automatically satisfied
in the lossless case. To be more specific, an $(n,R_{1},R_{2},R_{b},D_{1},D_{2},\Gamma,\epsilon)$
code is defined per Definition \ref{def_BC_lossless} with the difference
that there are two additional functions for the encoder, namely\begin{subequations}\label{eqn: en_recons-1}
\begin{align}
 & \psi_{1}\textrm{: }\mathcal{X}^{n}\rightarrow\mathcal{\hat{X}}_{1}^{n}\\
\textrm{and } & \psi_{2}\textrm{: }\mathcal{X}^{n}\rightarrow\mathcal{\hat{X}}_{2}^{n},
\end{align}
\end{subequations}which map the source sequence into the estimated
sequences at the encoder, namely $\psi_{1}(X^{n})$ and $\psi_{2}(X^{n})$,
respectively; and the CR requirements are imposed, \textit{\emph{i.e.}},\begin{subequations}\label{eqn: CR_req}
\begin{eqnarray}
\textrm{Pr}\left[\psi_{1}(X^{n})\neq\textrm{h}_{1}(M_{1},M_{b},Y^{n})\right] & \leq & \epsilon\\
\textrm{and Pr}\left[\psi_{2}(X^{n})\neq\textrm{h}_{2}(M_{2},M_{b},Z^{n})\right] & \leq & \epsilon,
\end{eqnarray}
\end{subequations}so that the encoder's estimates $\psi_{1}(\cdot$)
and $\psi_{2}(\cdot)$ are equal to the decoders' estimates (cf. (\ref{decoder1_BC})-(\ref{decoder2_BC}))
with high probability. 
\begin{prop}
\label{prop:RD_action_BC}The rate-distortion region $\mathcal{R\mbox{\ensuremath{(D_{1},D_{2},\Gamma)}}}$
for the cascade-broadcast source coding problem illustrated in Fig.
\ref{fig:fig3} under the CR constraint and the degradedness condition
(\ref{eq:degradedness}) is given by the union of all rate triples
$(R_{1},R_{2},R_{b})$ that satisfy the conditions\begin{subequations}\label{eqn: RD_action_BC}
\begin{eqnarray}
R_{b} & \geq & I(X;A)\label{eq:Ra}\\
R_{1}+R_{b} & \geq & I(X;A)+I(X;\hat{X}_{1},\hat{X}_{2}|A,Y)\\
R_{2}+R_{b} & \geq & I(X;A)+I(X;\hat{X}_{2}|A,Z)\\
\mbox{and}\mbox{ }R_{1}+R_{2}+R_{b} & \geq & I(X;A)+I(X;\hat{X}_{2}|A,Z)+I(X;\hat{X}_{1}|A,Y,\hat{X}_{2}),\label{eq:R1+R2+Ra}
\end{eqnarray}
\end{subequations}where the mutual information terms are evaluated
with respect to the joint pmf 
\begin{align}
p(x,y,z,a,\hat{x}_{1},\hat{x}_{2})=p(x)p(a|x)p(y|x,a)p(z|a,y)p(\hat{x}_{1},\hat{x}_{2}|x,a) & ,\label{eq:joint_BC}
\end{align}
such that the inequalities\begin{subequations}\label{eqn: action_cascade_BC}
\begin{eqnarray}
\ce{E}[d_{j}(X,\hat{X}_{j})] & \leq & D_{j}\textrm{, }\;\mbox{for }j=1,2,\label{eq:-1}\\
\mbox{and}\textrm{ }\ce{E}[\Lambda(A)] & \leq & \Gamma,
\end{eqnarray}
\end{subequations}are satisfied. \end{prop}
\begin{rem}
If either $R_{1}=0$ or $R_{b}=0$ and the side information $Y$ is
independent of the action $A$ given $X$$,$ i.e., $\ p(y|a,x)=p(y|x),$
the rate-distortion region $\mathcal{R}(D_{1},D_{2},\Gamma)$ of Proposition
\ref{prop:RD_action_BC} reduces to the one derived in \cite[Proposition 10]{Ahmadi_CR}. 
\end{rem}
The proof of the converse is provided in Appendix B. The achievability
follows similar to Proposition \ref{prop:RD_BC_lossless}. Specifically,
Node 1 first maps sequence $X^{n}$ into the action sequence $A^{n}$.
This mapping requires a codebook of rate $I(X;A)$. This rate has
to be conveyed over link $R_{b}$ by the definition of the problem
and is thus received by both Node 2 and Node 3. The source sequence
$X^{n}$ is mapped into the estimate $\hat{X}_{2}^{n}$ for Node 3
using a codebook of rate $I(X;\hat{X}_{2}|A)$ for each sequence $A^{n}$.
Communicating $\hat{X}_{2}^{n}$ to Node 2 requires rate $I(X;\hat{X}_{2}|A,Y)$
by the Wyner-Ziv theorem. We split this rate into two rates $r_{2b}$
and $r_{2d}$, such that the message corresponding to the first rate
is carried over the broadcast link of rate $R_{b}$ and the second
on the direct link of rate $R_{1}$. Note that Node 2 can thus recover
sequence $\hat{X}_{2}^{n}$. Communicating $\hat{X}_{2}^{n}$ to Node
3 requires rate $I(X;\hat{X}_{2}|A,Z)$ by the Wyner-Ziv theorem.
We split this rate into two rates $r_{0b}$ and $r_{0d}$. The message
corresponding to the rate $r_{0b}$ is send to Node 3 on the broadcast
link of the rate $R_{b}$ by Node 1, while the message of rate $r_{0d}$
is sent by Node 2 to Node 3. This way, Node 1 and Node 2 cooperate
to transmit $\hat{X}_{2}$ to Node 3. Finally, the source sequence
$X^{n}$ is mapped by Node 1 into the estimate $\hat{X}_{1}^{n}$
for Node 2 using a codebook of rate $I(X;\hat{X}_{1}|A,\hat{X}_{2})$
for each pair of sequences $(A^{n},\hat{X}_{2}^{n})$. Using the Wyner-Ziv
coding, this rate is reduced to $I(X;\hat{X}_{1}|A,Y,\hat{X}_{2})$
and split into two rates $r_{1b}$ and $r_{1d}$, which are sent through
links $R_{b}$ and $R_{1}$, respectively. As per discussion above,
the following inequalities have to be satisfied 
\begin{eqnarray*}
r_{0b}+r_{0d}+r_{2b} & \geq & I(X;\hat{X}_{2}|A,Z),\\
r_{2b}+r_{2d} & \geq & I(X;\hat{X}_{2}|A,Y),
\end{eqnarray*}
\begin{eqnarray*}
r_{1b}+r_{1d} & \geq & I(X;\hat{X}_{1}|A,Y,\hat{X}_{2}),\\
R_{1} & \geq & r_{1d}+r_{2d},\\
R_{2} & \geq & r_{0d},\\
\textrm{and }R_{b} & \geq & r_{1b}+r_{2b}+r_{0b}+I(X;A),
\end{eqnarray*}
Applying Fourier-Motzkin elimination \cite[Appendix C]{Elgammal}
to the inequalities above, the inequalities in (\ref{eqn: RD_action_BC})
are obtained.

\section{Adaptive Actions\label{sec:Adaptive-Actions}}

In this section, we assume that actions taken by the nodes are not
only a function of the message $M_{2}$ for the model of Fig. \ref{fig:fig2}
or $M_{b}$ for the models of Fig. \ref{fig:fig3} and Fig. \ref{fig:fig4},
respectively, but also a function of the past observed side information
samples. Following \cite{Chiru}, we refer to this case as the one
with \textit{adaptive actions}. Note that for the cascade-broadcast
problem, we consider the model in Fig. \ref{fig:fig4}, which differs
from the one in Fig. \ref{fig:fig3} considered thus far in that the
side information $Z$ is not available at Node 3. At this time, it
appears to be problematic to define adaptive actions in the presence
of two nodes that observe different side information sequences. For
the cascade model in Fig. \ref{fig:fig2}, a $(n,R_{1},R_{2},D_{1},D_{2},\Gamma)$
code is defined per Definition \ref{def_cascade} with the difference
that the action encoder (\ref{action_fun}) is modified to be 
\begin{equation}
\mathrm{\ell}\text{: }[1,2^{nR_{2}}]\times{\cal Z}^{i-1}\rightarrow\mathcal{A},\label{action_fun_adaptive}
\end{equation}
which maps the message $M_{2}$ and the past observed decoder side
information sequence $Z^{i-1}$ into the $i$th symbol of the action
sequence $A_{i}$. Moreover, for the cascade-broadcast model of Fig.
\ref{fig:fig4}, the {}``action\textquotedblright{}\ function (\ref{action_fun_BC})
in Definition \ref{def_BC_lossless} is modified as 
\begin{equation}
\mathrm{\ell}\text{: }[1,2^{nR_{b}}]\times{\cal Y}^{i-1}\rightarrow\mathcal{A},\label{action_fun_adaptive-1}
\end{equation}
which maps the message $M_{b}$ and the past observed decoder side
information sequence $Y^{i-1}$ into the $i$th symbol of the action
sequence $A_{i}$. 
\begin{prop}
\label{prop:RD_action_cascade_adaptive}The rate-distortion-cost region
$\mathcal{R\mbox{\ensuremath{(D_{1},D_{2},\Gamma)}}}$ for the cascade
source coding problem illustrated in Fig. \ref{fig:fig2} with adaptive
action-dependent side information is given by the rate region described
in Proposition \ref{prop:RD_action_cascade}. 
\begin{prop}
\label{prop:RD_action_BC_adaptive}The rate-distortion-cost region
$\mathcal{R\mbox{\ensuremath{(D_{1},D_{2},\Gamma)}}}$ for the cascade-broadcast
source coding problem under the CR illustrated in Fig. \ref{fig:fig4}
with adaptive action-dependent side information is given by the region
described in Proposition \ref{prop:RD_action_BC} by setting $Z=\emptyset$. 
\end{prop}
\end{prop}
\begin{rem}
The results above show that enabling adaptive actions does not increase
the achievable rate-distortion-cost region. These results generalize
the observations in \cite{Chiru} for the point-to-point setting,
wherein a similar conclusion is drawn.

To establish the propositions above, we only need to prove the converse.
The proofs for Proposition \ref{prop:RD_action_cascade_adaptive}
and Proposition \ref{prop:RD_action_BC_adaptive} are given in Appendix
A and B, respectively. 
\end{rem}

\section{Concluding Remarks}

In an increasing number of applications, communication networks are
expected to be able to convey not only data, but also information
about control for actuation over multiple hops. In this work, we have
tackled the analysis of a baseline communication model with three
nodes connected in a cascade with the possible presence of an additional
broadcast link. We have characterized the optimal trade-off between
rate, distortion and cost for actuation in a number of relevant cases
of interest. In general, the results point to the advantages of leveraging
a joint representation of data and control information in order to
utilize in the most efficient way the available communication links.
Specifically, in all the considered models, a layered coding strategy,
possibly coupled with rate splitting, has been proved to be optimal.
This strategy is such that the base layer has the double role of guiding
the actions of the downstream nodes and of providing a coarse description
of the source, similar to \cite{Permuter}. Moreover, it is shown
that this base compression layer should be designed in a way that
depends on the network topology and on the relative cost of activating
the different links.

\section{ACKNOWLEDGMENTS}

The work of O. Simeone is supported by the U.S. National Science Foundation
under grant CCF-0914899, and the work of U. Mitra by ONR N00014-09-1-0700,
NSF CCF-0917343 and DOT CA-26-7084-00.

\appendices{ }

\section*{Appendix A: Converse Proof for Proposition \ref{prop:RD_action_cascade}
and \ref{prop:RD_action_cascade_adaptive}}

Here, we prove the converse part of Proposition \ref{prop:RD_action_cascade_adaptive}.
Since the setting of Proposition \ref{prop:RD_action_cascade} is
more restrictive, as it does not allow for adaptive actions, the converse
proof for Proposition \ref{prop:RD_action_cascade} follows immediately.
For any $(n,R_{1},R_{2},D_{1}+\epsilon,D_{2}+\epsilon,\Gamma+\epsilon)$
code, we have
\begin{align}
nR_{1} & \geq H(M_{1})\nonumber \\
 & \geq H(M_{1}|Y^{n})\nonumber \\
 & \stackrel{(a)}{=}I(M_{1};X^{n},Z^{n}|Y^{n})\nonumber \\
 & =H(X^{n},Z^{n}|Y^{n})-H(X^{n},Z^{n}|M_{1},Y^{n})\nonumber \\
 & =H(X^{n}|Y^{n})+H(Z^{n}|X^{n},Y^{n})-H(Z^{n}|Y^{n},M_{1})-H(X^{n}|Z^{n},Y^{n},M_{1})\nonumber \\
 & \stackrel{(a,b)}{=}H(X^{n}|Y^{n})+H(Z^{n}|X^{n},Y^{n},M_{1},M_{2})-H(Z^{n}|Y^{n},M_{1},M_{2})-H(X^{n}|Z^{n},Y^{n},M_{1},M_{2})\nonumber \\
 & \stackrel{(c)}{=}H(X^{n}|Y^{n})-H(X^{n}|Z^{n},Y^{n},M_{1},M_{2},A^{n},\hat{X}_{1}^{n})\nonumber \\
 & +\sum_{i=1}^{n}H(Z_{i}|Z^{i-1},X^{n},Y^{n},M_{1},M_{2})-H(Z_{i}|Z^{i-1},Y^{n},M_{1},M_{2})\nonumber \\
 & \stackrel{(c,d)}{\geq}\sum_{i=1}^{n}(H(X_{i}|Y_{i})-H(X_{i}|X^{i-1},Y^{i},M_{2},A^{i},Z^{n},\hat{X}_{1i}))\nonumber \\
 & +\sum_{i=1}^{n}H(Z_{i}|Z^{i-1},X^{n},Y^{n},M_{1},M_{2},A_{i})-H(Z_{i}|Z^{i-1},Y^{n},M_{1},M_{2},A_{i})\nonumber \\
 & \stackrel{(e)}{=}\sum_{i=1}^{n}I(X_{i};\hat{X}_{1i},A_{i},U_{i}|Y_{i})+H(Z_{i}|Y_{i},A_{i})-H(Z_{i}|Y_{i},A_{i})\nonumber \\
 & =\sum_{i=1}^{n}I(X_{i};\hat{X}_{1i},A_{i},U_{i}|Y_{i}),\label{eq:conv1_end}
\end{align}
where $(a)$ follows since $M_{1}$ is a function of $(X^{n},Y^{n})$;
$(b)$ follows since $M_{2}$ is a function of $(M_{1},Y^{n})$; $(c)$
follows since $A_{i}$ is a function of $(M_{2},Z^{i-1})$ and since
$\hat{X}_{1}^{n}$ is a function of $(M_{1},Y^{n})$; $(d)$ follows
since conditioning decreases entropy and since $X^{n}$ and $Y^{n}$
are i.i.d.; and $(e)$ follows by defining $U_{i}=(M_{2},X^{i-1},Y^{i-1},A^{i-1},Z^{n\backslash i})$
and since $(Z^{i-1},X^{n},Y^{n\backslash i},M_{1},M_{2})\textrm{---}$
$(A_{i},Y_{i})\textrm{---}Z_{i}$ form a Markov chain by construction.
We also have
\begin{eqnarray*}
nR_{2} & \geq & H(M_{2})\\
 & = & I(M_{2};X^{n},Y^{n},Z^{n})
\end{eqnarray*}
\begin{eqnarray}
 & = & H(X^{n},Y^{n},Z^{n})-H(X^{n},Y^{n},Z^{n}|M_{2})\nonumber \\
 & = & H(X^{n},Y^{n})+H(Z^{n}|X^{n},Y^{n})-H(Z^{n}|M_{2})-H(X^{n},Y^{n}|M_{2},Z^{n})\nonumber \\
 & = & \overset{n}{\underset{i=1}{\sum}}H(X_{i},Y_{i})+H(Z_{i}|Z^{i-1},X^{n},Y^{n})-H(Z_{i}|Z^{i-1},M_{2})\nonumber \\
 &  & -H(X_{i},Y_{i}|X^{i-1},Y^{i-1},M_{2},Z^{n})\nonumber \\
 & \overset{(a)}{=} & \overset{n}{\underset{i=1}{\sum}}H(X_{i},Y_{i})\negmedspace+\negmedspace H(Z_{i}|Z^{i-1},X^{n},Y^{n},M_{2},A_{i})\negmedspace-\negmedspace H(Z_{i}|Z^{i-1},M_{2},A_{i})\negmedspace\nonumber \\
 &  & -\negmedspace H(X_{i},Y_{i}|X^{i-1},Y^{i-1},M_{2},Z^{n},A^{i})\nonumber \\
 & \overset{(b)}{\geq} & \overset{n}{\underset{i=1}{\sum}}H(X_{i},Y_{i})+H(Z_{i}|X_{i},Y_{i},A_{i})-H(Z_{i}|A_{i})-H(X_{i},Y_{i}|U_{i},A_{i},Z_{i}),\label{eq:conv2_end}
\end{eqnarray}
where (\textit{a}) follows because $M_{2}$ is a function of $(M_{1},Y^{n})$
and thus of $(X^{n},Y^{n})$ and because $A^{i}$ is a function of
$(M_{2},Z^{i-1})$ and ($b$) follows since conditioning decreases
entropy, since the Markov chain relationship $Z_{i}\textrm{---}(X_{i},Y_{i},A_{i})\textrm{---}$
$(X^{n\backslash i},Y^{n\backslash i},M_{2})$ holds and by using
the definition of $U_{i}$.

Defining $Q$ to be a random variable uniformly distributed over $[1,n]$
and independent of all the other random variables and with $X\overset{\triangle}{=}X_{Q}$,
$Y\overset{\triangle}{=}Y_{Q}$, $Z\overset{\triangle}{=}Z_{Q}$,
$A\overset{\triangle}{=}A_{Q}$, $\hat{X}_{1}\overset{\triangle}{=}\hat{X}_{1Q}$,
$\hat{X}_{2}\overset{\triangle}{=}\hat{X}_{2Q}$ and $U\overset{\triangle}{=}(U_{Q},Q),$
from (\ref{eq:conv1_end}) we have 
\begin{eqnarray*}
nR_{1} & \geq & I(X;\hat{X}_{1},A,U|Y,Q)\overset{(a)}{\geq}H(X|Y)-H(X|\hat{X}_{1},A,U,Y)=I(X;\hat{X}_{1},A,U|Y),
\end{eqnarray*}
where in ($a$) we have used the fact that $(X^{n},Y^{n})$ are i.i.d
and conditioning reduces entropy. Moreover, from (\ref{eq:conv2_end})
we have 
\begin{eqnarray*}
nR_{2} & \geq & H(X,Y|Q)+H(Z|X,Y,A,Q)-H(Z|A,Q)-H(X,Y|U,A,Z,Q)\\
 & \overset{(a)}{\geq} & H(XY)+H(Z|X,Y,A)-H(Z|A)-H(X,Y|U,A,Z)\\
 & = & I(XY;U,A,Z)-I(Z;X,Y|A)\\
 & = & I(XY;A)+I(X,Y;U|A,Z),
\end{eqnarray*}
where ($a$) follows since $(X^{n},Y^{n})$ are i.i.d, since conditioning
decreases entropy, by the definition of $U$ and by the problem definition.
We note that the defined random variables factorize as (\ref{eq:joint})
since we have the Markov chain relationship $X$---$(A,Y)$---$Z$
by the problem definition and that $\hat{X}_{2}$ is a function $\textrm{f}(U,Z)$
of $U$ and $Z$ by the definition of $U$. Moreover, from the cost
and distortion constraints (\ref{action cost})-(\ref{dist const}),
we have\begin{subequations}\label{eqn: action_cascade_const-1} 
\begin{align}
D_{j}+\epsilon & \geq\textrm{\ensuremath{\frac{1}{n}\sum_{i=1}^{n}}E}[d_{j}(X_{i},\hat{X}_{ji})]=\textrm{E}[d_{j}(X,\hat{X}_{j})],\textrm{ for }j=1,2,\label{eq:dist1-1}\\
\textrm{and }\Gamma+\epsilon & \geq\frac{1}{n}\sum_{i=1}^{n}\textrm{E}\left[\Lambda(A_{i})\right]=\textrm{E}\left[\Lambda(A)\right].
\end{align}
\end{subequations}

To bound the cardinality of auxiliary random variable $U$, we fix
$p(z|y,a)$ and factorize the joint pmf $p(x,y,z,a,u,\hat{x}_{1})$
as 
\begin{eqnarray*}
p(x,y,z,a,u,\hat{x}_{1}) & = & p(u)p(\hat{x}_{1},a,x,y|u)p(z|y,a).
\end{eqnarray*}
Therefore, for fixed $p(z|y,a)$, the quantities (\ref{eq:R1})-(\ref{eq:action_bound})
can be expressed in terms of integrals given by $\int g_{j}(p(\hat{x}_{1},a,x,y|u))dF(u)$,
for $j=1,...,|\mathcal{X}||\mathcal{Y}||\mathcal{A}|+3$, of functions
$g_{j}(\cdot)$ that are continuous on the space of probabilities
over alphabet $|\mathcal{X}|\text{\texttimes}|\mathcal{Y}|\text{\texttimes}|\mathcal{A}|\text{\texttimes}|\hat{\mathcal{X}}_{1}|$.
Specifically, we have $g_{j}$ for $j=1,...,|\mathcal{X}||\mathcal{Y}||\mathcal{A}|-1$,
given by the pmf $p(a,x,y)$ for all values of $x\in\mathcal{X}$,
$y\in\mathcal{Y}$ and $a\in\mathcal{A}$, (except one), $g_{|\mathcal{X}||\mathcal{Y}||\mathcal{A}|}=H(X|A,Y,\hat{X}_{1},U=u)$,
$g_{|\mathcal{X}||\mathcal{Y}||\mathcal{A}|+1}=H(X,Y|A,Z,U=u)$, and
$g_{|\mathcal{X}||\mathcal{Y}||\mathcal{A}|+1+j}=\textrm{E}[d_{j}(X,\hat{X}_{j})|U=u],$
for $j=1,2$. The proof in concluded by invoking the Fenchel--Eggleston--Caratheodory
theorem \cite[Appendix C]{Elgammal}.

\section*{Appendix B: Proof of Proposition \ref{prop:RD_action_BC}}

Here, we prove the converse parts of Proposition \ref{prop:RD_action_BC}
and Proposition \ref{prop:RD_action_BC_adaptive}. We start by proving
Proposition \ref{prop:RD_action_BC}. The proof of Proposition \ref{prop:RD_action_BC_adaptive}
will follow by setting $Z=\emptyset$, and noting that in the proof
below the action $A_{i}$ can be made to be a function of $Y^{i-1}$,
in addition to being a function of $M_{b}$, without modifying any
steps of the proof. By the CR requirements (\ref{eqn: CR_req}), we
first observe that for any $(n,R_{1},R_{2},R_{b},D_{1}+\epsilon,D_{2}+\epsilon,\Gamma+\epsilon)$
code, we have the Fano inequalities\begin{subequations}\label{eqn: Fano}
\begin{eqnarray}
H(\psi_{1}(X^{n})|\textrm{h}_{1}(M_{1},M_{b},Y^{n})) & \leq & n\delta(\epsilon),\label{eq:Fano1}\\
\textrm{and }H(\psi_{2}(X^{n})|\textrm{h}_{2}(M_{2},M_{b},Z^{n})) & \leq & n\delta(\epsilon),\label{eq:Fano2}
\end{eqnarray}
\end{subequations}where $\delta(\epsilon)$ denotes any function
such that $\delta(\epsilon)\rightarrow0$ if $\epsilon\rightarrow0$.
Next, we have
\begin{eqnarray*}
 & nR_{b} & \geq H(M_{b})\\
 &  & \stackrel{(a)}{=}I(M_{b};X^{n},Y^{n})
\end{eqnarray*}
\begin{align}
 & =H(X^{n},Y^{n})-H(X^{n},Y^{n}|M_{b})\nonumber \\
 & \stackrel{(a)}{=}H(X^{n})+H(Y^{n}|X^{n},M_{b})-H(X^{n},Y^{n}|M_{b})\nonumber \\
 & \stackrel{(b)}{=}\sum_{i=1}^{n}H(X_{i})+H(Y_{i}|Y^{i-1},X^{n},M_{b},A_{i})-H(X_{i},Y_{i}|X^{i-1},Y^{i-1},M_{b},A_{i})\nonumber \\
 & =\sum_{i=1}^{n}H(X_{i})+H(Y_{i}|Y^{i-1},X^{n},M_{b},A_{i})-H(X_{i}|X^{i-1},Y^{i-1},M_{b},A_{i})\nonumber \\
 & -H(Y_{i}|X^{i},Y^{i-1},M_{b},A_{i})\nonumber \\
 & \stackrel{(c)}{=}\sum_{i=1}^{n}H(X_{i})+H(Y_{i}|X_{i},A_{i})-H(X_{i}|X^{i-1},Y^{i-1},M_{b},A_{i})-H(Y_{i}|X_{i},A_{i})\nonumber \\
 & \stackrel{(d)}{\geq}\sum_{i=1}^{n}H(X_{i})-H(X_{i}|A_{i}),\label{eq:conv_Rb}
\end{align}

where $(a)$ follows since $M_{b}$ is a function of $X^{n}$; $(b)$
follows since $A_{i}$ is a function of $M_{b}$ and since $X^{n}$
is i.i.d.; $(c)$ follows since $(Y^{i-1},X^{n\backslash i},M_{b})\textrm{---}(A_{i},X_{i})\textrm{---}Y_{i}$
forms a Markov chain by problem definition; and $(d)$ follows conditioning
reduces entropy. In the following, for simplicity of notation, we
write $\textrm{h}_{1},\textrm{h}_{2},\psi_{1},\psi_{2}$ for the values
of corresponding functions in Sec. \ref{sub:CR}. Next, We can also
write
\begin{eqnarray*}
n(R_{1}+R_{b}) & \geq & H(M_{1},M_{b})\\
 & \overset{(a)}{=} & I(M_{1},M_{b};X^{n},Y^{n},Z^{n})\\
 & = & H(X^{n},Y^{n},Z^{n})-H(X^{n},Y^{n},Z^{n}|M_{1},M_{b})\\
 & = & H(X^{n})+H(Y^{n},Z^{n}|X^{n})-H(Y^{n},Z^{n}|M_{1},M_{b})-H(X^{n}|Y^{n},Z^{n},M_{1},M_{b})\\
 & \overset{(b)}{=} & H(X^{n})+H(Y^{n},Z^{n}|X^{n},M_{b})-H(Y^{n}|M_{1},M_{b})\\
 &  & -H(Z^{n}|M_{1},M_{b},Y^{n},A^{n})-H(X^{n}|Y^{n},Z^{n},M_{1},M_{b},M_{2},A^{n})\\
 & \overset{(b,c)}{=} & \overset{n}{\underset{i=1}{\sum}}H(X_{i})+H(Y_{i},Z_{i}|X_{i},A_{i})-H(Y_{i}|Y^{i-1},M_{1},M_{b},A_{i})\\
 &  & -H(Z_{i}|Z^{i-1},M_{1},M_{b},Y^{n},A^{n})-H(X_{i}|X^{i-1},Y^{n},Z^{n},M_{1},M_{b},A^{n},M_{2},\textrm{h}_{1},\textrm{h}_{2})\\
 & \overset{(d)}{\geq} & \overset{n}{\underset{i=1}{\sum}}H(X_{i})\negmedspace+\negmedspace H(Y_{i}|X_{i},A_{i})+\negmedspace H(Z_{i}|Y_{i},A_{i})-H(Y_{i}|A_{i})-H(Z_{i}|Y_{i},A_{i})\\
 &  & -H(X_{i}|Y_{i},A_{i},\textrm{h}_{1},\textrm{h}_{2})\negmedspace
\end{eqnarray*}
\begin{eqnarray}
 & = & \overset{n}{\underset{i=1}{\sum}}I(X_{i};Y_{i},A_{i},\textrm{h}_{1},\textrm{h}_{2})-I(Y_{i};X_{i}|A_{i})\nonumber \\
 & = & \overset{n}{\underset{i=1}{\sum}}I(X_{i};Y_{i},A_{i},\textrm{h}_{1},\textrm{h}_{2},\psi_{1},\psi_{2})-I(X_{i};\psi_{1},\psi_{2}|Y_{i},A_{i},\textrm{h}_{1},\textrm{h}_{2})-I(Y_{i};X_{i}|A_{i})\nonumber \\
 & \overset{(e)}{\geq} & \overset{n}{\underset{i=1}{\sum}}I(X_{i};Y_{i},A_{i},\psi_{1},\psi_{2})-H(\psi_{1},\psi_{2}|Y_{i},A_{i},\textrm{h}_{1},\textrm{h}_{2})\nonumber \\
 &  & +H(\psi_{1},\psi_{2}|Y_{i},A_{i},\textrm{h}_{1},\textrm{h}_{2},X_{i})-I(Y_{i};X_{i}|A_{i})\nonumber \\
 & \overset{(f)}{\geq} & \overset{n}{\underset{i=1}{\sum}}I(X_{i};Y_{i},A_{i},\psi_{1},\psi_{2})-I(Y_{i};X_{i}|A_{i})+n\delta(\epsilon)\nonumber \\
 & = & \overset{n}{\underset{i=1}{\sum}}I(X_{i};A_{i})+I(X_{i};\psi_{1},\psi_{2}|Y_{i},A_{i})+n\delta(\epsilon),\label{eq:conv_R1+Rb}
\end{eqnarray}
where ($a$) follows because $(M_{1},M_{b})$ is a function of $X^{n}$;
($b$) follows because $M_{b}$ is a function of $X^{n}$, $A^{n}$
is a function of $M_{b}$ and $M_{2}$ is a function of $(M_{1},M_{b},Y^{n})$;
($c$) follows since $H(Y^{n},Z^{n}|X^{n},M_{b})=\sum_{i=1}^{n}H(Y_{i},Z_{i}|$
$Y^{i-1},Z^{i-1},X^{n},M_{b},A_{i})=\sum_{i=1}^{n}H(Y_{i},Z_{i}|X_{i},A_{i})$
and since $\textrm{h}_{1}$ and $\textrm{h}_{2}$ are functions of
$(M_{1},M_{b},Y^{n})$ and $(M_{2},M_{b},Z^{n})$, respectively and
because $(Y_{i},Z_{i})\textrm{---}(X_{i},A_{i})\textrm{---}$ $(X^{n\backslash i},Y^{i-1},Z^{i-1},M_{b})$
forms a Markov chain; ($d$) follows since conditioning reduces entropy,
since side information VM follows $p(y^{n},z^{n}|a^{n},x^{n})$$=\prod_{i=1}^{n}p_{Y|A,X}(y_{i}|a_{i},x_{i})$
$p_{Z|A,Y}(z_{i}|a_{i},y_{i})$ from (\ref{eq:degradedness}) and
because $Z_{i}\textrm{---}(Y_{i},A_{i})\textrm{---}$ $(Y^{n\backslash i},Z^{i-1},M_{1},M_{b})$
forms a Markov chain; ($e$) follows by the chain rule for mutual
information and the fact that mutual information is non-negative;
and ($f$) follows by the Fano inequality (\ref{eqn: Fano}) and because
entropy is non-negative. We can also write
\begin{eqnarray*}
n(R_{2}+R_{b}) & \geq & H(M_{2},M_{b})\\
 & \stackrel{(a)}{=} & I(M_{2},M_{b};X^{n},Y^{n},Z^{n})\\
 & = & H(X^{n},Y^{n},Z^{n})-H(X^{n},Y^{n},Z^{n}|M_{2},M_{b})\\
 & \stackrel{(a)}{=} & H(X^{n})+H(Y^{n},Z^{n}|X^{n},M_{b})-H(Z^{n}|M_{2},M_{b})-H(X^{n},Y^{n}|Z^{n},M_{2},M_{b})\\
 & \stackrel{(b)}{=} & \sum_{i=1}^{n}H(X_{i})+H(Y_{i},Z_{i}|Y^{i-1},Z^{i-1},X^{n},M_{b},A_{i})-H(Z_{i}|Z^{i-1},M_{2},M_{b},A_{i})\\
 & - & H(X_{i},Y_{i}|X^{i-1},Y^{i-1},M_{2},M_{b},Z^{n},A_{i})\\
 & = & \sum_{i=1}^{n}H(X_{i},Y_{i})-H(Y_{i}|X_{i})+H(Y_{i},Z_{i}|Y^{i-1},Z^{i-1},X^{n},M_{b},A_{i})\\
 & - & H(Z_{i}|Z^{i-1},M_{2},M_{b},A_{i})-H(X_{i},Y_{i}|X^{i-1},Y^{i-1},M_{2},M_{b},Z^{n},A_{i})
\end{eqnarray*}
\begin{align}
 & \stackrel{(c)}{=}\sum_{i=1}^{n}H(X_{i},Y_{i})-H(Y_{i}|X_{i})+H(Y_{i}|X_{i},A_{i})+H(Z_{i}|A_{i},Y_{i},X_{i})\nonumber \\
 & -H(Z_{i}|Z^{i-1},M_{2},M_{b},A_{i})-H(X_{i},Y_{i}|X^{i-1},Y^{i-1},M_{2},M_{b},Z^{n},A_{i})\nonumber \\
 & \stackrel{(d)}{=}\sum_{i=1}^{n}H(X_{i},Y_{i})-I(Y_{i};A_{i}|X_{i})+H(Z_{i}|A_{i},Y_{i},X_{i})-H(Z_{i}|Z^{i-1},M_{2},M_{b},A_{i})\nonumber \\
 & -H(X_{i},Y_{i}|X^{i-1},Y^{i-1},M_{2},M_{b},\textrm{h}_{2},Z^{n},A_{i})\nonumber \\
 & \stackrel{(e)}{\geq}\sum_{i=1}^{n}H(X_{i},Y_{i})+I(X_{i};A_{i})-I(Y_{i},X_{i};A_{i})+H(Z_{i}|A_{i},Y_{i},X_{i})\nonumber \\
 & -H(Z_{i}|A_{i})-H(X_{i},Y_{i}|\textrm{h}_{2},A_{i},Z_{i})\nonumber \\
 & =\sum_{i=1}^{n}I(X_{i},Y_{i};\textrm{h}_{2},A_{i},Z_{i},\psi_{2i})-I(X_{i},Y_{i};\psi_{2i}|\textrm{h}_{2},A_{i},Z_{i})+I(X_{i};A_{i})\nonumber \\
 & -I(Y_{i},X_{i};A_{i})-I(X_{i},Y_{i};Z_{i}|A_{i})\nonumber \\
 & \geq\sum_{i=1}^{n}I(X_{i},Y_{i};A_{i},Z_{i},\psi_{2i})-H(\psi_{2i}|\textrm{h}_{2},A_{i},Z_{i})+H(\psi_{2i}|\textrm{h}_{2},A_{i},X_{i},Y_{i},Z_{i})\nonumber \\
 & +I(X_{i};A_{i})-I(X_{i},Y_{i};Z_{i},A_{i})\nonumber \\
 & \stackrel{(f)}{\geq}\sum_{i=1}^{n}I(X_{i};A_{i})+I(X_{i},Y_{i};\psi_{2i}|A_{i},Z_{i})+n\delta(\epsilon),\label{eq:conv_R2+Rb}
\end{align}
where $(a)$ follows since $M_{b}$ is a function of $X^{n}$ and
because $M_{2}$ is a function of $(M_{1},M_{b},Y^{n})$ and thus
of $(X^{n},Y^{n})$; $(b)$ follows since $A_{i}$ is a function of
$M_{b}$ and since $X^{n}$ is i.i.d.; $(c)$ follows since $(Y_{i},Z_{i})\textrm{---}(X_{i},A_{i})\textrm{---}$
$(X^{n\backslash i},Y^{i-1},Z^{i-1},M_{b})$ forms a Markov chain
and since $p(y^{n},z^{n}|a^{n},x^{n})$ $=\prod_{i=1}^{n}p_{Y|A,X}(y_{i}|a_{i},x_{i})$$p_{Z|A,Y}(z_{i}|a_{i},y_{i})$;
$(d)$ follows since $\textrm{h}_{2}$ is a function of $(M_{2},M_{b},Z^{n})$;
$(e)$ follows since conditioning reduces entropy; and $(f)$ follows
since entropy is non-negative and using the Fanos inequality. Moreover,
with the definition $M=(M_{1},M_{2},M_{b})$, we have the chain of
inequalities
\begin{align*}
n(R_{1}+R_{2}+R_{b}) & \geq H(M)\\
 & \stackrel{(a)}{=}I(M;X^{n},Y^{n},Z^{n})\\
 & =H(X^{n},Y^{n},Z^{n})-H(X^{n},Y^{n},Z^{n}|M)\\
 & \stackrel{(a)}{=}H(X^{n})+H(Y^{n},Z^{n}|X^{n},M_{b})-H(X^{n},Y^{n},Z^{n}|M)
\end{align*}
\begin{align}
 & =I(X^{n};A^{n})+H(Y^{n},Z^{n}|X^{n},M_{b})-H(Y^{n},Z^{n}|M)\nonumber \\
 & -H(X^{n}|Y^{n},Z^{n},M)+H(X^{n}|A^{n})\nonumber \\
 & =I(X^{n};A^{n})+H(Y^{n},Z^{n}|X^{n},M_{b})-H(Y^{n},Z^{n}|M)+I(X^{n};Y^{n},Z^{n},M|A^{n})\nonumber \\
 & =I(X^{n};A^{n})+I(M;X^{n}|Y^{n},A^{n},Z^{n})+H(Y^{n},Z^{n}|X^{n},M_{b})\nonumber \\
 & -H(Y^{n},Z^{n}|M)+I(X^{n};Y^{n},Z^{n}|A^{n})\nonumber \\
 & \stackrel{(b)}{=}H(X^{n})-H(X^{n}|A^{n})+H(X^{n}|Y^{n},A^{n},Z^{n})-H(X^{n}|Y^{n},A^{n},Z^{n},M)\nonumber \\
 & -H(Y^{n},Z^{n}|M)+H(Y^{n},Z^{n}|A^{n})\nonumber \\
 & =H(X^{n})-H(X^{n}|A^{n})+H(X^{n},Y^{n},Z^{n}|A^{n})-H(X^{n}|Y^{n},A^{n},Z^{n},M)\nonumber \\
 & -H(Y^{n},Z^{n}|M)\nonumber \\
 & =H(X^{n})+H(Y^{n},Z^{n}|A^{n},X^{n})-H(X^{n}|Y^{n},A^{n},Z^{n},M)-H(Y^{n},Z^{n}|M)\nonumber \\
 & \stackrel{(c)}{=}\sum_{i=1}^{n}H(X_{i})+H(Y_{i}|A_{i},X_{i})+H(Z_{i}|A_{i},Y_{i})-H(X_{i}|X^{i-1},Y^{n},A^{n},Z^{n},M)\nonumber \\
 & -H(Z_{i}|Z^{i-1},M,A_{i})-H(Y_{i}|Y^{i-1},Z^{n},M,A_{i})\nonumber \\
 & \stackrel{(d)}{=}\sum_{i=1}^{n}H(X_{i})\negmedspace+\negmedspace H(Y_{i}|A_{i},X_{i})\negmedspace+\negmedspace H(Z_{i}|A_{i},Y_{i})\negmedspace-\negmedspace H(X_{i}|X^{i-1},\negmedspace Y^{n},A^{n},Z^{n},M,\textrm{h}_{1},\textrm{h}_{2})\nonumber \\
 & -H(Z_{i}|Z^{i-1},M,A_{i})-H(Y_{i}|Y^{i-1},Z^{n},M,A_{i},\textrm{h}_{2})\nonumber \\
 & \geq\sum_{i=1}^{n}H(X_{i})+H(Y_{i}|A_{i},X_{i})+H(Z_{i}|A_{i},Y_{i})-H(X_{i}|Y_{i},A_{i},\textrm{h}_{1},\textrm{h}_{2})\nonumber \\
 & -H(Z_{i}|A_{i})-H(Y_{i}|Z_{i},A_{i},\textrm{h}_{2})\nonumber \\
 & \stackrel{(e)}{\geq}I(X_{i};A_{i},Y_{i},\psi_{1},\psi_{2})+H(Y_{i}|A_{i},X_{i})+H(Z_{i}|A_{i},Y_{i})\nonumber \\
 & -H(Z_{i}|A_{i})-H(Y_{i}|Z_{i},A_{i},\psi_{2})-n\delta(\epsilon),\label{eq:conv_end-1}
\end{align}
where $(a)$ follows since $(M_{1},M_{b})$ is a function of $X^{n}$
and $M_{2}$ is a function of $(M_{1},M_{b},Y^{n})$; $(b)$ follows
since $H(Y^{n},Z^{n}|X^{n},M_{b})=\sum_{i=1}^{n}H(Y_{i},Z_{i}|$ $Y^{i-1},Z^{i-1},X^{n},M_{b},A_{i})=\sum_{i=1}^{n}H(Y_{i},Z_{i}|$
$X_{i},A_{i})=H(Y^{n},Z^{n}|X^{n},A^{n})$; $(c)$ follows since $A_{i}$
is a function of $M_{b}$; $(d)$ follows since $\textrm{h}_{1},\textrm{h}_{2}$
are functions of $(M,Y^{n})$ and $(M,Z^{n})$, respectively; and
$(e)$ follows since entropy is non-negative and by Fano's inequality.
Next, from (\ref{eq:conv_end-1}) we have

\textbf{
\begin{align}
n(R_{1}+R_{2}+R_{b}) & \geq I(X_{i};A_{i},Y_{i},\psi_{1},\psi_{2})+H(Y_{i}|A_{i},X_{i})+H(Z_{i}|A_{i},Y_{i})-H(Z_{i}|A_{i})\nonumber \\
 & -H(Y_{i},Z_{i}|A_{i},\psi_{2})+H(Z_{i}|A_{i},\psi_{2})-n\delta(\epsilon)\nonumber \\
 & =I(X_{i};A_{i},Y_{i},\psi_{1},\psi_{2})+H(Y_{i}|A_{i},X_{i})-H(Z_{i}|A_{i})-H(Y_{i}|A_{i},\psi_{2})\nonumber \\
 & +H(Z_{i}|A_{i},\psi_{2})-n\delta(\epsilon)\nonumber \\
 & =I(X_{i};A_{i},Y_{i},\psi_{1},\psi_{2})-I(X_{i};Y_{i}|A_{i},\psi_{2})-I(Z_{i};\psi_{2}|A_{i})-n\delta(\epsilon)\nonumber \\
 & \stackrel{(a)}{=}I(X_{i};A_{i},Y_{i},\psi_{1},\psi_{2})-I(X_{i};Y_{i}|A_{i},\psi_{2})-I(Y_{i};A_{i}|X_{i})-I(Z_{i};Y_{i}|A_{i})\nonumber \\
 & +I(Y_{i};A_{i},\psi_{2}|X_{i})+I(Z_{i};Y_{i}|\psi_{2},A_{i})-n\delta(\epsilon)\nonumber \\
 & \stackrel{(b)}{=}I(X_{i};A_{i},Y_{i},\psi_{1},\psi_{2})-I(X_{i};Y_{i}|A_{i},\psi_{2})+I(X_{i};A_{i})-I(Y_{i},X_{i};A_{i})\nonumber \\
 & -I(Z_{i};X_{i},Y_{i}|A_{i})\negmedspace+\negmedspace I(X_{i},Y_{i};A_{i},\psi_{2})\negmedspace+\negmedspace I(Z_{i};X_{i},Y_{i}|\psi_{2},A_{i})-\negmedspace I(X_{i};A_{i},\psi_{2})\negmedspace-\negmedspace n\delta(\epsilon)\nonumber \\
 & =I(X_{i};A_{i})+I(X_{i};A_{i},Y_{i},\psi_{1},\psi_{2})+I(X_{i},Y_{i};A_{i},\psi_{2},Z_{i})-I(A_{i},Z_{i};X_{i},Y_{i})\nonumber \\
 & -I(X_{i};Y_{i},A_{i},\psi_{2})-n\delta(\epsilon)\nonumber \\
 & =I(X_{i};A_{i})\negmedspace+\negmedspace I(X_{i};A_{i},Y_{i},\psi_{1},\psi_{2})\negmedspace+\negmedspace I(X_{i},Y_{i};\psi_{2}|A_{i},Z_{i})\negmedspace-\negmedspace I(X_{i};Y_{i},A_{i},\psi_{2})\negmedspace-\negmedspace n\delta(\epsilon)\nonumber \\
 & =I(X_{i};A_{i})+I(X_{i},Y_{i};\psi_{2}|A_{i},Z_{i})+I(X_{i};\psi_{1}|A_{i},Y_{i},\psi_{2})-n\delta(\epsilon),\label{eq:conv_end}
\end{align}
}where $(a)$ is true since 
\begin{eqnarray*}
 &  & I(Y_{i};A_{i}|X_{i})+I(Z_{i};Y_{i}|A_{i})-I(Y_{i};A_{i},\psi_{2}|X_{i})-I(Z_{i};Y_{i}|\psi_{2},A_{i})\\
 &  & =H(Y_{i}|X_{i})-H(Y_{i}|X_{i},A_{i})+H(Z_{i}|A_{i})-H(Z_{i}|A_{i},Y_{i})-H(Y_{i}|X_{i})+H(Y_{i}|X_{i},A_{i})\\
 &  & -H(Z_{i}|\psi_{2},A_{i})+H(Z_{i}|A_{i},Y_{i})\\
 &  & =H(Z_{i}|A_{i})-H(Z_{i}|\psi_{2},A_{i});
\end{eqnarray*}
$(b)$ follows because $I(Z_{i};X_{i},Y_{i}|A_{i})=I(Z_{i};Y_{i}|A_{i})$
and $I(Z_{i};X_{i},Y_{i}|A_{i},\psi_{2})=I(Z_{i};Y_{i}|A_{i},\psi_{2})$.

Next, define $\hat{X}_{ji}=\psi_{ji}(X^{n})$ for $j=1,2$ and $i=1,2,...,n$
and let $Q$ be a random variable uniformly distributed over $[1,n]$
and independent of all the other random variables and with $X\overset{\triangle}{=}X_{Q}$,
$Y\overset{\triangle}{=}Y_{Q}$, $A\overset{\triangle}{=}A_{Q}$,
from (\ref{eq:conv_Rb}), we have 
\begin{align*}
nR_{b} & \overset{}{\geq}H(X|Q)-H(X|A,Q)\overset{(a)}{\geq}H(X)-H(X|A)=I(X;A),
\end{align*}
where ($a$) follows since $X^{n}$ is i.i.d. and since conditioning
decreases entropy. Next, from (\ref{eq:conv_R1+Rb}), we have$ $
\begin{eqnarray*}
n(R_{1}+R_{b}) & \overset{}{\geq} & I(X;A|Q)+I(X;\hat{X}_{1},\hat{X}_{2}|Y,A,Q)\\
 & \overset{(a)}{\geq} & I(X;A)+I(X;\hat{X}_{1},\hat{X}_{2}|Y,A),
\end{eqnarray*}
where ($a$) follows since $X^{n}$ is i.i.d., since conditioning
decreases entropy and by the problem definition. From (\ref{eq:conv_R2+Rb}),
we also have
\begin{eqnarray*}
n(R_{2}+R_{b}) & \overset{}{\geq} & I(X;A|Q)+I(X,Y;\hat{X}_{2}|A,Z,Q)\\
 & \overset{(a)}{\geq} & I(X;A)+H(X,Y|A,Z,Q)-H(X,Y|A,Z,\hat{X}_{2})\\
 & \overset{(b)}{=} & I(X;A)+H(Y|A,Z)+H(X|A,Y,Z)-H(X,Y|A,Z,\hat{X}_{2})\\
 & = & I(X;A)+I(X,Y;\hat{X}_{2}|A,Z)\\
 & \geq & I(X;A)+I(X;\hat{X}_{2}|A,Z)
\end{eqnarray*}
where ($a$) follows since $X^{n}$ is i.i.d. and by conditioning
reduces entropy; and ($b$) follows by the problem definition. Finally,
from (\ref{eq:conv_end}), we have 
\begin{align}
n(R_{1}+R_{2}+R_{b}) & \overset{}{\geq}I(X,A|Q)+I(X,Y;\hat{X}_{2}|A,Z,Q)+I(X;\hat{X}_{1}|A,Y,\hat{X}_{2},Q)\nonumber \\
 & \overset{(a)}{\geq}I(X,A)+H(X,Y|A,Z,Q)-H(X,Y|A,Z,\hat{X}_{2})+I(X;\hat{X}_{1}|A,Y,\hat{X}_{2})\nonumber \\
 & \overset{(b)}{=}I(X;A)\negmedspace+\negmedspace H(Y|A,Z)\negmedspace+\negmedspace H(X|A,Y,Z)\negmedspace-\negmedspace H(X,Y|A,Z,\hat{X}_{2})\negmedspace+\negmedspace I(X;\hat{X}_{1}|A,Y,\hat{X}_{2})\nonumber \\
 & =I(X;A)+I(X,Y;\hat{X}_{2}|A,Z)+I(X;\hat{X}_{1}|A,Y,\hat{X}_{2})\nonumber \\
 & \geq I(X;A)+I(X;\hat{X}_{2}|A,Z)+I(X;\hat{X}_{1}|A,Y,\hat{X}_{2})
\end{align}
where ($a$) follows since $X^{n}$ is i.i.d, since conditioning decreases
entropy, and by the problem definition; and ($b$) follows by the
problem definition. From cost constraint (\ref{action cost}), we
have 
\begin{align}
\Gamma+\epsilon & \geq\frac{1}{n}\sum_{i=1}^{n}\textrm{E}\left[\Lambda(A_{i})\right]=\textrm{E}\left[\Lambda(A)\right].
\end{align}

Moreover, let $\mathcal{B}$ be the event $\mathcal{B}=\{\left(\psi_{1}\textrm{(}X^{n})\neq h_{1}(M_{1},M_{b},Y^{n})\right)\wedge\left(\psi_{2}\textrm{(}X^{n})\neq h_{2}(M_{2},M_{b})\right)\}$.
Using the CR requirement (\ref{eqn: CR_req}), we have $\textrm{Pr}(\mathcal{B})\leq\epsilon$.
For $j=1,2$, we have 
\begin{eqnarray}
\textrm{E}\left[d(X_{j},\hat{X}_{j})\right] & = & \frac{1}{n}\sum_{i=1}^{n}\textrm{E}\left[d(X_{ji},\hat{X}_{ji})\right]\nonumber \\
 & = & \frac{1}{n}\sum_{i=1}^{n}\textrm{E}\negmedspace\left[d(X_{ji},\hat{X}_{ji})\Big|\mathcal{B}\right]\negmedspace\textrm{Pr}(\mathcal{B})\negmedspace+\negmedspace\frac{1}{n}\sum_{i=1}^{n}\textrm{E}\negmedspace\left[d(X_{ji},\hat{X}_{ji})\Big|\mathcal{B}^{c}\right]\negmedspace\textrm{Pr}(\mathcal{B}^{c})\nonumber \\
 & \negmedspace\negmedspace\overset{(a)}{\leq}\negmedspace\negmedspace & \frac{1}{n}\sum_{i=1}^{n}\textrm{E}\left[d(X_{ji},\hat{X}_{ji})\Big|\mathcal{B}^{c}\right]\textrm{Pr}(\mathcal{B}^{c})+\epsilon D_{max}\nonumber \\
 & \negmedspace\negmedspace\overset{(b)}{\leq}\negmedspace\negmedspace & \frac{1}{n}\sum_{i=1}^{n}\textrm{E}\left[d(X_{ji},h_{ji})\right]+\epsilon D_{max}\nonumber \\
 & \negmedspace\negmedspace\overset{(c)}{\leq}\negmedspace\negmedspace & D_{j}+\epsilon D_{max},
\end{eqnarray}
where ($a$) follows using the fact that $\textrm{Pr}(\mathcal{B})\leq\epsilon$
and that the distortion is upper bounded by $D_{max}$; ($b$) follows
by the definition of $\hat{X}_{ji}$ and $\mathcal{B}$; and ($c$)
follows by (\ref{dist const}).


\begin{thebibliography}{1}

\bibitem{Permuter}H. Permuter and T. Weissman, {}``Source coding
with a side information {}``vending machine\textquotedblright{},\textquotedblright{}\ \textit{IEEE
Trans. Inf. Theory}, vol. 57, pp. 4530--4544, Jul 2011.

\bibitem{Ravi}R. Tandon, S. Mohajer, and H. V. Poor, {}``Cascade
source coding with erased side information,\textquotedblright{} in
\textit{Proc. IEEE Symp. Inform. Theory}, St. Petersburg, Russia,
Aug. 2011.

\bibitem{Vasudevan}D. Vasudevan, C. Tian, and S. N. Diggavi, {}``Lossy
source coding for a cascade communication system with side-informations,''
In \textit{Proc. 44th Annual Allerton Conference on Communications,
Control and Computing}, Monticello, IL, September 2006.

\bibitem{Chia}Y. K. Chia, H. Permuter and T. Weissman, {}``Cascade,
triangular and two way source coding with degraded side information
at the second user,\textquotedblright{}\ http://arxiv.org/abs/1010.3726.

\bibitem{Weissman_multi}Y. Chia, H. Asnani, and T. Weissman, \textquotedbl{}Multi-terminal
source coding with action dependent side information,\textquotedbl{}
in \emph{Proc. IEEE International Symposium on Information Theory}
(ISIT 2011), July 31-Aug. 5, Saint Petersburg, Russia, 2011.

\bibitem{Ahmadi_ISIT'11}B. Ahmadi and O. Simeone, \textquotedbl{}Robust
coding for lossy computing with receiver-side observation costs,\textquotedbl{}
in \emph{Proc. IEEE International Symposium on Information Theory}
(ISIT 2011), July 31-Aug. 5, Saint Petersburg, Russia, 2011.

\bibitem{HB}C. Heegard and T. Berger, \textquotedbl{}Rate distortion
when side information may be absent,\textquotedbl{} \emph{IEEE Trans.
Inf. Theory}, vol. 31, no. 6, pp. 727-734, Nov. 1985.

\bibitem{Kaspi}A. Kaspi, \textquotedbl{}Rate-distortion when side-information
may be present at the decoder,\textquotedbl{} \emph{IEEE Trans. Inf.
Theory}, vol. 40, no. 6, pp. 2031--2034, Nov. 1994.

\bibitem{Ahmadi_DSC}B. Ahmadi and O. Simeone, {}``Distributed and
cascade lossy source coding with a side information \textquotedbl{}Vending
Machine\textquotedbl{},\textquotedblright{} http://arxiv.org/abs/1109.6665.

\bibitem{Berger-Yeung}T. Berger and R. Yeung, {}``Multiterminal
source encoding with one distortion criterion,\textquotedblright{}\ \textit{IEEE
Trans. Inform. Theory}, vol. 35, pp. 228--236, Mar 1989.

\bibitem{Compression with actions}L. Zhao, Y. K. Chia, and T. Weissman,
{}``Compression with actions,'' in Allerton conference on communications,
control and computing, Monticello, Illinois, September 2011.

\bibitem{Ahmadi_CR}B. Ahmadi, R. Tandon, O. Simeone, and H. V. Poor,
{}``Heegard-Berger and cascade source coding problems with common
reconstruction constraints,\textquotedblright{} arXiv:1112.1762v3,
2011.

\bibitem{Steinberg} Y. Steinberg, {}``Coding and common reconstruction,\textquotedblright{}\ \textit{IEEE
Trans. Inform. Theory}, vol. 55, no. 11, 2009.

\bibitem{Weiss-Elgam}T. Weissman and A. El Gamal, {}``Source coding
with limited-look-ahead side information at the decoder,\textquotedblright{}\textit{
IEEE Trans. Inf. Theory}, vol. 52, no. 12, pp. 5218--5239, Dec. 2006.

\bibitem{Chiru}C. Choudhuri and U. Mitra, {}``How useful is adaptive
action?,\textquotedblright{} Submitted to Globecom 2012.

\bibitem{Elgammal} A. El Gamal and Y. Kim, \textit{Network Information
Theory}, Cambridge University Press, Dec 2011.

\end{thebibliography}
\end{document}